\documentclass[aps,prd,nofootinbib,floatfix,superscriptaddress,preprint,eqsecnum,tightenlines,longbibliography]{revtex4-1}

\usepackage[utf8]{inputenc}
\usepackage{amsfonts}
\usepackage{amsmath}
\usepackage{amssymb}
\usepackage{bm}
\usepackage{graphicx}
\usepackage[pdftex]{color}
\usepackage[sort&compress]{natbib}
\usepackage[colorlinks=true,linkcolor=blue,filecolor=blue,urlcolor=blue,citecolor=blue,pdftex,plainpages=false]{hyperref}

\begin{document}

\newcommand\bnl{Physics Department, Brookhaven National Laboratory, Upton, NY 11973, USA}
\newcommand\arizona{Department of Physics, University of Arizona, Tucson, AZ 85721, USA}
\newcommand\rbrc{RIKEN BNL Research Center, Brookhaven National Laboratory, Upton, NY 11973, USA}
\newcommand\uconn{Physics Department, University of Connecticut, Storrs, CT 06269-3046, USA}
\newcommand\cu{Physics Department, Columbia University, New York, NY 10027, USA}
\newcommand\cern{CERN, Physics Department, 1211 Geneva 23, Switzerland}
\newcommand\fnal{Theoretical Physics Department, Fermi National Accelerator Laboratory, \\ Batavia, IL 60510-5011, USA}
\newcommand\uiuc{Department of Physics, University of Illinois Urbana-Champaign, \\ Urbana, IL 61801-3003, USA}
\newcommand\syracuse{Department of Physics, Syracuse University, Syracuse, NY 13244, USA}
\newcommand\colorado{Department of Physics, University of Colorado, Boulder, CO 80309, USA}
\newcommand\uwash{Physics Department, University of Washington, Seattle WA 98195-1560, USA}

\title{Opportunities for lattice QCD in quark and lepton flavor physics}
\author{Christoph Lehner}
\email{Editor, \tt clehner@quark.phy.bnl.gov}
\affiliation{\bnl}
\author{Stefan Meinel}
\email{Editor, \tt smeinel@email.arizona.edu}
\affiliation{\arizona}
\affiliation{\rbrc}
\author{Tom Blum}
\affiliation{\uconn}
\author{Norman H.~Christ}
\affiliation{\cu}
\author{Aida~X.~El-Khadra}
\affiliation{\uiuc}
\author{Maxwell T.~Hansen}
\affiliation{\cern}
\author{Andreas S.~Kronfeld}
\affiliation{\fnal}
\author{Jack~Laiho}
\affiliation{\syracuse}
\author{Ethan T.~Neil}
\affiliation{\colorado}
\affiliation{\rbrc}
\author{Stephen R.~Sharpe}
\affiliation{\uwash}
\author{Ruth S.~Van de Water}
\affiliation{\fnal}
\collaboration{USQCD Collaboration}
\noaffiliation

\date{November 19, 2019}
\begin{abstract}
This document is one of a series of whitepapers from the USQCD collaboration.  Here,
we discuss opportunities for lattice QCD in quark and lepton flavor
physics. New data generated at Belle II, LHCb, BES III, NA62, KOTO,
and Fermilab E989, combined with precise calculations of the relevant
hadronic physics, may reveal what lies beyond the Standard Model. We
outline a path toward improvements of the precision of existing
lattice-QCD calculations and discuss groundbreaking new methods that
allow lattice QCD to access new observables.
\end{abstract}

\maketitle

\section*{Executive Summary}

\label{sec:ES}

In 2018, the USQCD collaboration’s Executive Committee organized several subcommittees to recognize future opportunities
and formulate possible goals for lattice field theory calculations in several physics areas.
The conclusions of these studies, along with community input, are presented in seven
whitepapers~\cite{Bazavov:2018qcd,Brower:2018qcd,Davoudi:2018qcd,Detmold:2018qcd,Joo:2018qcd,Kronfeld:2018qcd}.
This whitepaper covers the role of lattice QCD in quark and lepton flavor physics.

Flavor physics provides a window to look beyond the Standard Model of elementary particles, in many cases reaching farther than direct searches at
high-energy colliders.  With experiments that are dramatically
improving in precision now and in the coming years, flavor physics may
very well reveal where the Standard Model fails, and point us toward a
more fundamental theory.  Concrete opportunities arise from new data
generated at Belle II, LHCb, BES III, NA62, KOTO, and Fermilab E989.
These experiments, paired with improvements in theory, will shed new
light on existing tensions between theory and experiment, such as those in
the flavor-changing neutral-current $b\to s \mu^+ \mu^-$ transitions and in the muon anomalous magnetic
moment.

Most of the observables in flavor physics involve hadrons, and their
theoretical description therefore requires nonperturbative
calculations in QCD.  In many cases, the lack of precision of
theoretical predictions limits the power of the
experiments to constrain the Standard Model and to search for new physics.
The only systematically improvable method for nonperturbative calculations in QCD is lattice gauge theory, which
has now reached a high level of maturity. In this whitepaper,
we discuss opportunities for lattice QCD to fully exploit the upcoming
and existing experimental results in flavor physics.
We outline a path toward improvements of the precision of existing
lattice-QCD calculations, as well as groundbreaking new methods that
allow lattice QCD to access new observables and therefore tap the
potential of a larger variety of experimental measurements.

In quark flavor physics, examples of established observables where improvements in precision over previous lattice calculations will have a big impact
are the $B_s$--$\bar{B}_s$ mixing matrix elements and the $B\to\pi$ form
factors, while examples of more complicated observables that can now be calculated on the lattice are decays
with two-hadron final states such as $K\to\pi\pi$ and $B\to K\pi \ell^+ \ell^-$,
and non-local electroweak processes such as the long-distance contributions to the CP-violation amplitude
$\varepsilon_K$ and rare kaon decays. There are also new proposals to compute inclusive processes, which involve a sum over
arbitrary hadronic states, on the lattice. In lepton flavor physics, lattice-QCD calculations of
the hadronic vacuum polarization function for the muon anomalous magnetic moment are well established, and
the goal for the next few years is to match or better the high level of precision provided by dispersive extractions
from experimental data. A more complicated observable is the hadronic light-by-light contribution to the muon anomalous magnetic moment,
where recent improvements in the methodology have made possible a complete first-principles calculation.

The opportunities outlined in this whitepaper build upon USQCD's existing, highly successful program in flavor physics.
In particular, lattice QCD calculations of some of the important quark flavor observables have reached a level of
precision where there are now a number of quantities for which lattice uncertainties are commensurate with
(or smaller than) experimental uncertainties. In addition, for some key observables, breakthrough lattice results
(while not yet at commensurate precision) are pointing to an emerging tension between experiment and Standard-Model theory.
Taken all together, these results are considered to be among the flagship results obtained in lattice QCD and
have a big impact on the corresponding phenomenology in the Standard Model and beyond.

The hardware resources of the USQCD collaboration have been a crucial component of this successful program,
as they allowed USQCD researchers to develop, test, and refine new methods and other innovations, in addition
to carrying out the needed computations on all but the most demanding ensembles.  The USQCD hardware resources
will continue to be important to develop and test new methods needed for computations of more challenging observables
involving multi-hadron intermediate or final states, or the more complicated sub-leading corrections needed
to meet the precision needs of the experimental program. However, the availability of allocations on leadership-class
facilities will also continue to play an important role in facilitating further improvements by allowing
lattice calculations for mature projects on the most demanding ensembles.

\newpage

\tableofcontents

\section{Introduction}
\label{sec:Intro}
With the discovery of the Higgs boson in 2012, the last elementary particle of the Standard Model was found, but several unexplained
phenomena and theoretical arguments suggest that we still do not have a complete theory.
The violation of CP symmetry in the quark sector that originates from the complex phase of the Cabibbo-Kobayashi-Maskawa quark
mixing matrix is many orders of magnitude too small to explain the matter-antimatter asymmetry of the universe.
The Standard Model does not provide a particle suitable as the dominant constituent of the observed dark matter.
The mechanism giving neutrinos their tiny masses is still unknown, and the patterns and hierarchies of the masses and couplings of
the many ``elementary'' particles remain puzzling.

Flavor physics in the quark sector has a proven track record of discoveries of new fundamental physics: the unexpectedly low
frequency of neutral kaon oscillations led to the prediction of the charm quark, the observation of CP violation in the same sector
demanded the existence of the third-generation bottom and top quarks, and
measurements of $B^0$-$\bar{B}^0$ mixing indicated a large value of the top-quark mass before colliders were able to produce top quarks directly \cite{Schwarzschild:1987sm}.
Similarly, in the lepton sector, the anomalous magnetic moment of electrons and muons has played a pivotal role in advancing our
understanding of relativity and quantum field theory.
Theory and experiment for these moments can now be compared at a precision of approximately one part in five million for the muon
and one part in eight billion for the electron.
In both the quark and lepton sector, indirect searches for new fundamental physics using low-energy flavor observables can probe
energy scales beyond those directly accessible in particle collisions, and can also probe very weakly coupled new light particles.
These indirect searches powerfully complement direct searches looking for the production of new elementary particles.

\begin{figure}
    \includegraphics[width=0.5\linewidth]{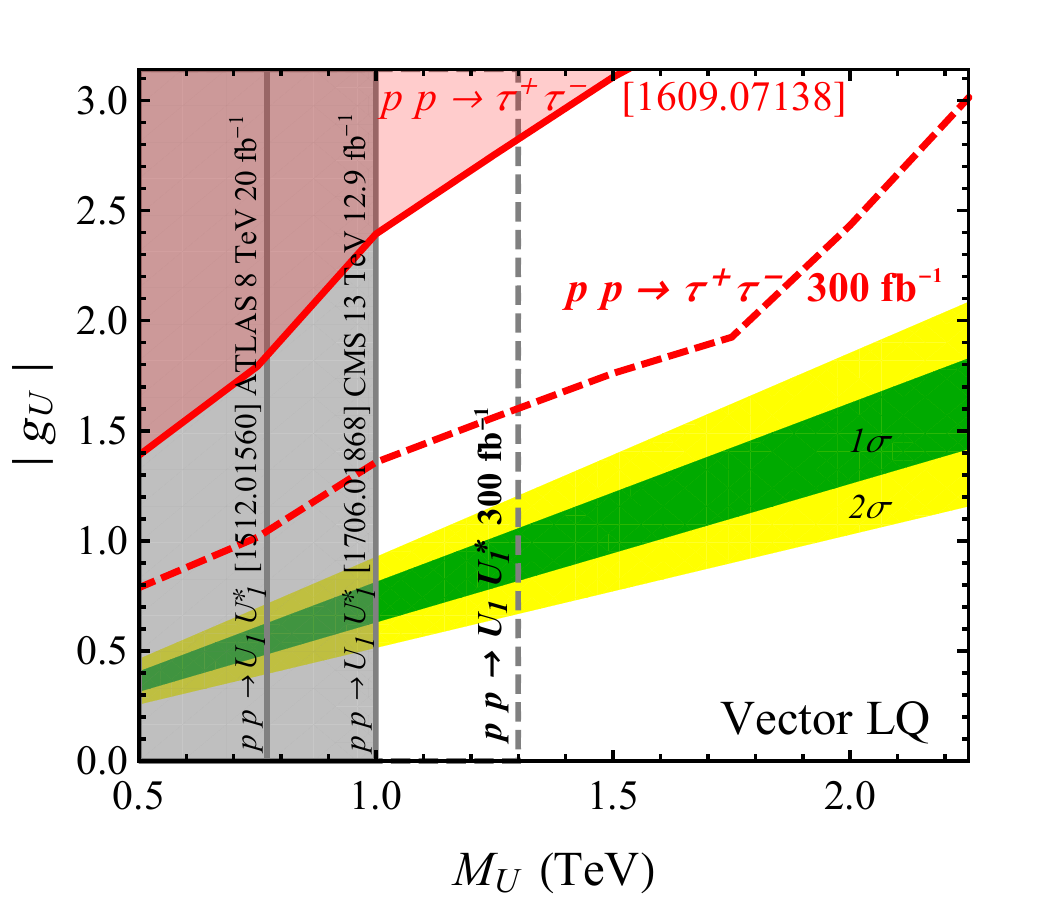}
    \caption{Indirect searches for new physics in the flavor sector and searches at high transverse momentum are
        complementary.
        Here one sees the constraints on the mass and coupling of a
        vector leptoquark $U$ \cite{Buttazzo:2017ixm}.
        The regions labeled ``$1\sigma$'' and ``$2\sigma$'' can explain the currently observed anomalies in two classes of $B$-meson
        decays, observed primarily by LHCb, Babar, and Belle (see Fig.~\protect\ref{fig:Banomalies}).
        The shaded regions in the left and top-left parts of the plot are excluded by searches at high transverse momentum with the
        ATLAS and CMS experiments. From Ref.~\cite{Buttazzo:2017ixm}.}
    \label{fig:LHC-U1}
\end{figure}

There are already a number of intriguing tensions between experimental data and Standard-Model predictions.
These include the tension in the muon anomalous magnetic moment \cite{Davier:2017zfy,Keshavarzi:2018mgv,Blum:2018mom} and hints for
violation of lepton flavor universality in decays of bottom quarks observed by BaBar, Belle, and LHCb \cite{Archilli:2017xmu,Ciezarek:2017yzh,Bifani:2018zmi}.
Further investigating these tensions through improvements in experiment and theory, and also searching for possible heavy mediators
beyond the Standard Model in high-energy collisions, are high priorities.
An example for the complementarity between the flavor-physics observables and searches at high energy is shown in
Fig.~\ref{fig:LHC-U1}, which contrasts the reach of LHC experiments with the properties of a leptoquark that would explain the 
flavor anomalies~\cite{Buttazzo:2017ixm}.

Since the strong force is present almost everywhere, lattice QCD calculations are essential to make the connection between the
experimental data and the fundamental short-distance processes.
Many important observables in flavor physics, such as decay rates or oscillation frequencies, depend on hadronic matrix elements,
which must be computed nonperturbatively from first principles.
The use of gauge-field ensembles at the physical pion mass is now standard in flavor physics, and with ultrafine lattices it has
become possible to treat even the bottom quark relativistically.
These developments have recently enabled calculations of the $D$, $D_s$, $B$, and $B_s$ decay constants by the Fermilab Lattice and MILC Collaborations
with subpercent precision \cite{Bazavov:2017lyh}. The quark masses $m_u$, $m_d$, $m_s$, $m_c$, and $m_b$ have been
determined with similar precision \cite{Bazavov:2018omf,Lytle:2018evc}, which in fact inspired theoretical work on 
quark masses with possible wider applicability~\cite{Komijani:2017vep,Brambilla:2017hcq}.
In addition, lattice technology has advanced significantly, such that complex calculations that have previously appeared out of
reach for current computing hardware are now feasible.
A prominent example of this is the \emph{ab-initio} calculation of the hadronic light-by-light contribution to the muon anomalous
magnetic moment by the RBC collaboration \cite{Blum:2016lnc}.

In this whitepaper, we outline future opportunities for lattice QCD in flavor physics.
We identify observables where improvements in the precision are needed to match experiments, as well as observables that have not
previously been calculated in lattice QCD but are now within reach.
Section \ref{sec:QF} covers quark flavor physics, while Sec.~\ref{sec:LF} discusses charged-lepton flavor physics.
We begin each section with a summary of the experimental motivation, before discussing the relevant lattice calculations.

\section{Quark flavor physics}
\label{sec:QF}
\subsection{Experimental motivation}
\label{sec:quarkexp}

The question whether the Cabibbo-Kobayashi-Maskawa (CKM) mechanism of the Standard Model completely describes flavor-changing
interactions of quarks, and is the only source of CP violation in this sector, lies at the heart of quark flavor physics.
Major experimental efforts are underway to constrain the elements of the CKM matrix using many different
processes, to test whether all these processes can indeed be described by a common unitary matrix.
These experiments also search for processes that are very rare or forbidden in the Standard Model, but could receive observable
contributions from possible new fundamental interactions.
In many cases, lattice QCD calculations are needed to make the connection between the fundamental parameters of interest and the
experimentally observed processes.

The Large Hadron Collider has just completed its Run 2 and has delivered vast amounts of proton-proton collisions to the LHCb, CMS,
ATLAS, and ALICE experiments.
The number of collisions in the LHCb experiment to date corresponds to the production of approximately one trillion ($10^{12}$)
pairs of bottom quarks and antiquarks~\cite{Aaij:2016avz}, which then formed all possible types of bottom mesons and baryons.
The number of charm-quark pairs produced is another order of magnitude larger~\cite{Aaij:2015bpa}, which has recently allowed
the first observation of CP violation in charm decays \cite{Aaij:2019kcg}.
Following the LHCb upgrade~\cite{Bediaga:2012uyd}, another two trillion $b\bar{b}$ pairs will be produced in LHC Run 3 (scheduled
for 2021--2023) and another five trillion in LHC Run 4 (scheduled for 2026--2029)~\cite{Albrecht:2017odf}.
The LHCb upgrade II is planned in 2030 in preparation for the high-luminosity LHC era~\cite{Bediaga:2018lhg}.

In the earlier Babar and Belle experiments, only about one billion ($10^9$) $b\bar{b}$ pairs of bottom quarks and
antiquarks were produced, but the production mechanism $e^+ e^- \to \Upsilon(4S) \to B \bar{B}$ used there provides additional
kinematic constraints that are strongly advantageous in particular for decays with undetected neutrinos.
The new Belle II experiment, which started running in 2018, also uses this production
mechanism, with a rate up to 40 times higher than in Belle~\cite{Kou:2018nap}.
The Belle II experiment is expected to take data for a total of approximately 50 billion $B\bar{B}$ pairs during
2018--2025~\cite{Kou:2018nap}.
This data set will allow precise measurements of many decay modes that are not easily accessible with a hadron collider.

The BESIII experiment is similar to Babar and Belle (II) but has a lower beam energy and focuses on charm
quarks~\cite{Asner:2008nq}.
BESIII has already performed many precise measurements involving charm mesons and also charm baryons and is expected to continue
taking data at various beam energies for several years~\cite{BESIII}.
Belle II will also study charm-meson decays and is expected to substantially exceed the statistics of BESIII~\cite{Kou:2018nap}.

In the strange-meson sector, several key processes related to CP violation and rare decays were already measured by past
experiments (including KTeV~\cite{KTeV}) far more precisely than even today's best
theoretical predictions, and improved calculations using lattice QCD will have a big impact.
There are two new experiments dedicated to decays of strange mesons.
NA62 aims to measure the branching fraction of the rare kaon decay $K^+ \to \pi^+ \nu \bar{\nu}$ with approximately 10\%
uncertainty~\cite{NA62:2312430}, following up on Brookhaven E949~\cite{Artamonov:2008qb}.
A first candidate event was observed in 2018~\cite{CortinaGil:2018fkc}.
KOTO focuses on the similar rare kaon decay $K_L\to\pi^0\nu\bar{\nu}$~\cite{2012PTEP.2012bB006Y}, and first results were also
published in 2018~\cite{Ahn:2018mvc}.

The most commonly considered test of CKM unitarity is that of the orthogonality of the first and third rows.
This orthogonality condition becomes a sum of three complex numbers which, when plotted in the complex plane, should form a
triangle.
After normalizing the bottom side of the triangle to unit length, the real and imaginary parts of the apex of the triangle are given
by the Wolfenstein parameters $\bar{\rho}$ and $\bar{\eta}$.
\begin{figure}
    \includegraphics[width=0.4\linewidth]{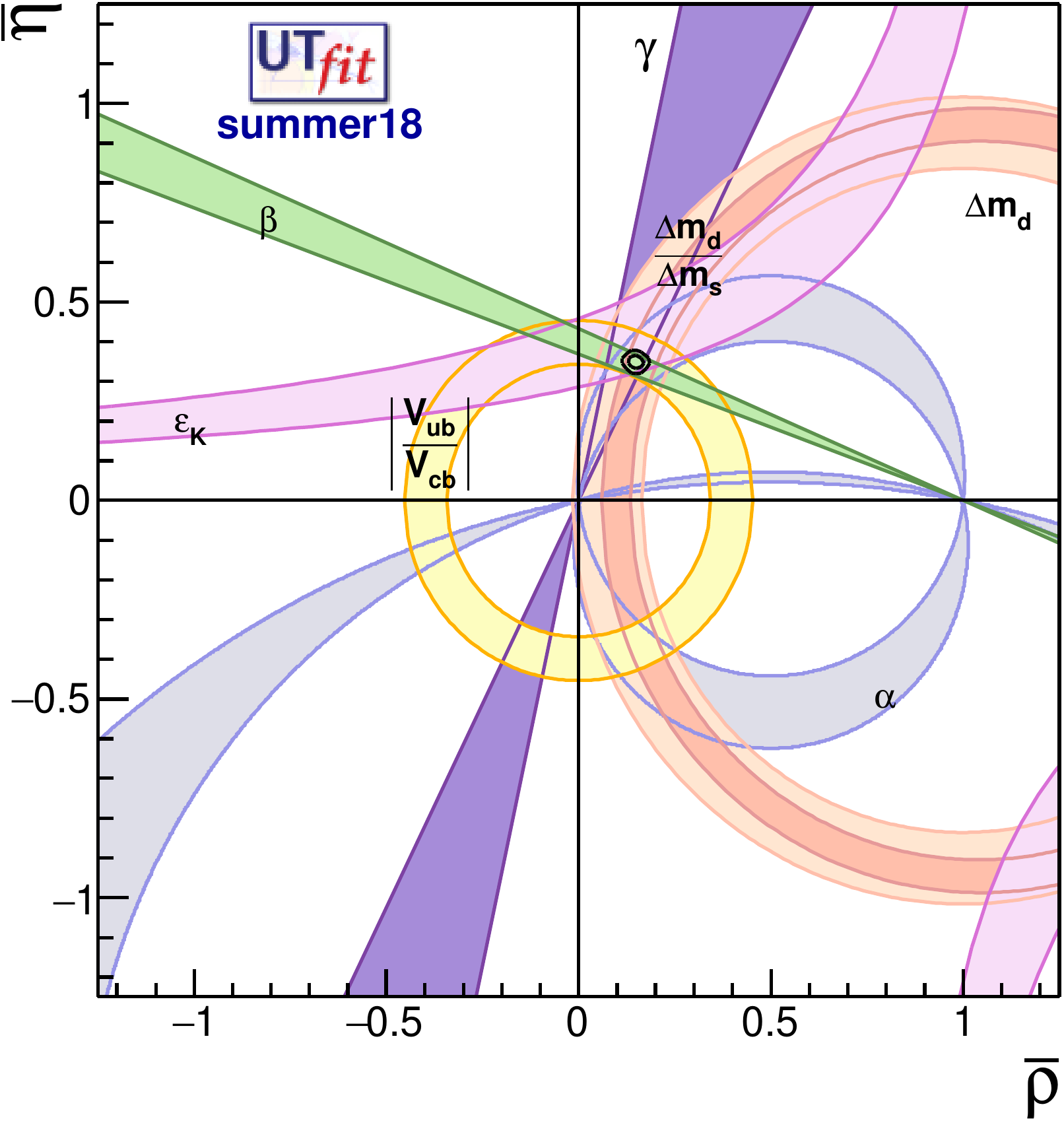}
    \caption{Constraints on two of the four independent parameters of the Cabibbo-Kobayashi-Maskawa
        quark mixing matrix, as of summer 2018~\cite{UTfit}.
        These constraints are obtained by combining experimental measurements of decay rates and other observables with calculations
        in the Standard Model, in many cases using lattice QCD.
        The constraints presently all overlap in the small region outlined in black that corresponds to the apex of the unitarity
        triangle.
        With more precise lattice QCD calculations and future experimental data, inconsistencies due to physics beyond the Standard
        Model might be revealed.
        In fact, two classes of $B$ meson decay processes that are not included here already show tensions with the Standard Model
        (see Fig.~\protect\ref{fig:Banomalies}). From Ref.~\cite{UTfit}.}
    \label{fig:unitaritytriangle}
\end{figure}
\begin{figure}
    \includegraphics[width=0.4\linewidth]{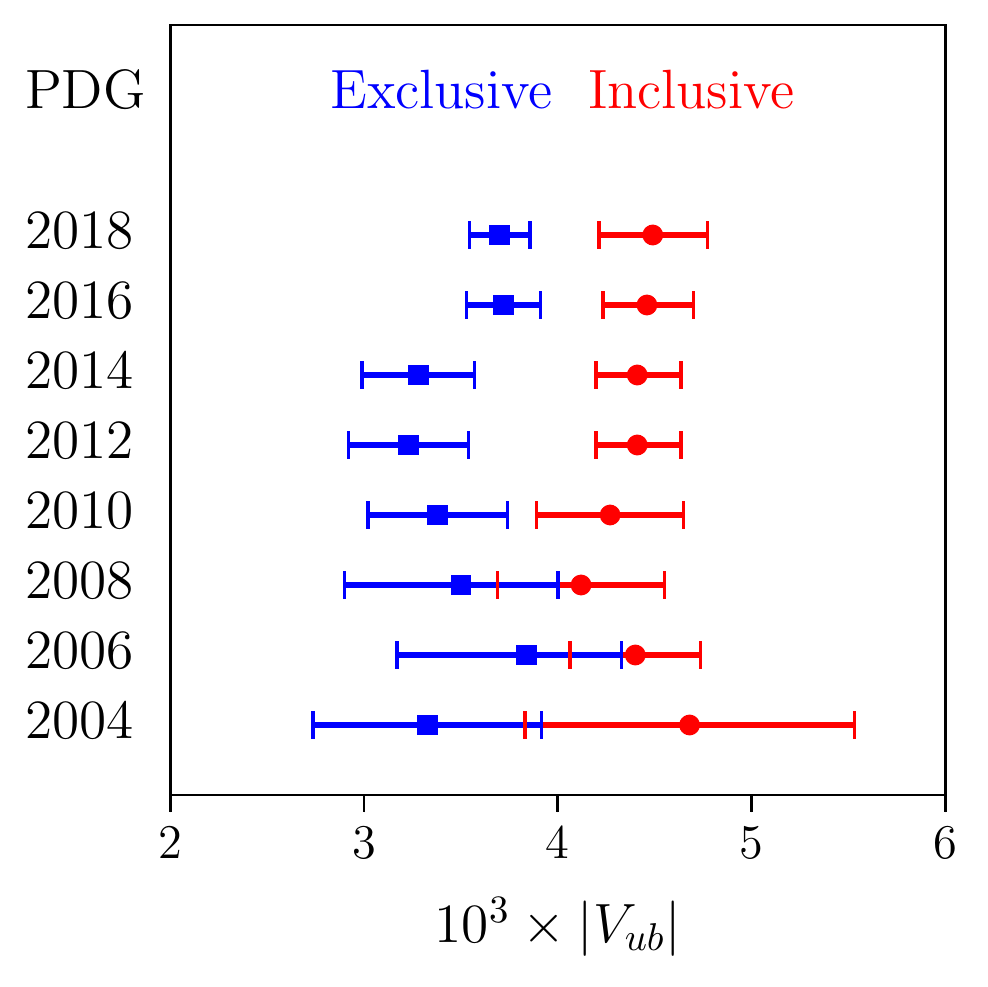}
    \caption{The smallest and least well known element of the CKM quark mixing matrix is $V_{ub}$.
    As illustrated here with the history of the values reported by the Particle Data Group~\cite{Tanabashi:2018oca}, there is a
    long-standing tension between two different methods of extracting $|V_{ub}|$: ``exclusive'', which combines measurements of the
    $B\to \pi \ell \bar{\nu}$ decay rate with lattice QCD calculations, and ``inclusive'', where the sum of the decay rates to all
    possible up-flavored hadrons in the final state is used.} 
    \label{fig:Vubhistory}
\end{figure}
The present constraints on these parameters from several different observables are shown as the shaded regions in
Fig.~\ref{fig:unitaritytriangle}.
The constraints presently all overlap in the small region outlined in black that corresponds to the apex of the unitarity triangle.
With more precise lattice QCD calculations and future experimental data, inconsistencies due to physics beyond the Standard Model
might be revealed.
For example, reducing the width of the yellow circle constraining the left side of the triangle could result in an inconsistency
with the precisely measured angle $\beta$ opposite to that side.
The width of the yellow circle is presently dominated by the uncertainty in $|V_{ub}|$, which is extracted primarily from
semileptonic $B$ meson decays.
For $|V_{ub}|$, there is a long-standing tension between determinations from $B \to \pi \ell \bar{\nu}$, where the calculation is
done using lattice QCD and determinations from an inclusive sample of final states (see Fig.~\ref{fig:Vubhistory}).
The related CKM matrix element $|V_{cb}|$ enters in the normalization of the triangle and is currently also the dominant source of
uncertainty in the constraint labeled ``$\epsilon_K$''~\cite{Bailey:2018feb} in Fig.~\ref{fig:unitaritytriangle}.
The Belle II experiment will provide $1\%$-level uncertainties for the decay rates used to determine $|V_{ub}|$ and $|V_{cb}|$, and
it is imperative to match these uncertainties with future lattice QCD calculations.
The LHCb experiment can also determine $|V_{ub}/V_{cb}|$, in particular using $\Lambda_b$ baryon~\cite{Aaij:2015bfa} and $B_s$-meson decays.
As lattice calculations of $K\to\pi\pi$ mature, the resulting constraints will allow for an additional horizontal band to be added
to the CKM unitarity plot.

\begin{figure}
    \parbox{0.49\linewidth}{
    \includegraphics[width=\linewidth]{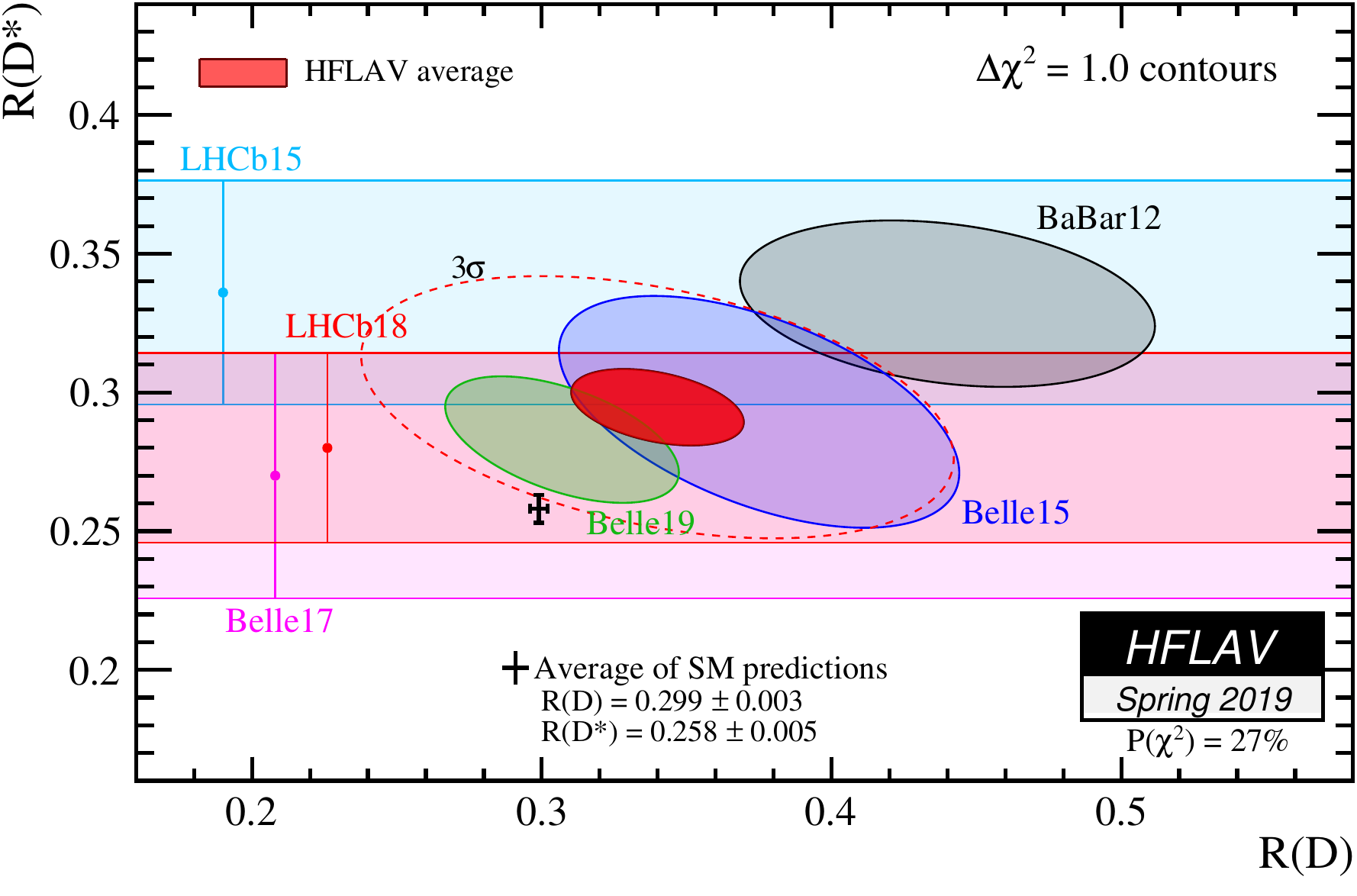}
    }
    \hfill
    \parbox{0.49\linewidth}{    
    \includegraphics[width=\linewidth]{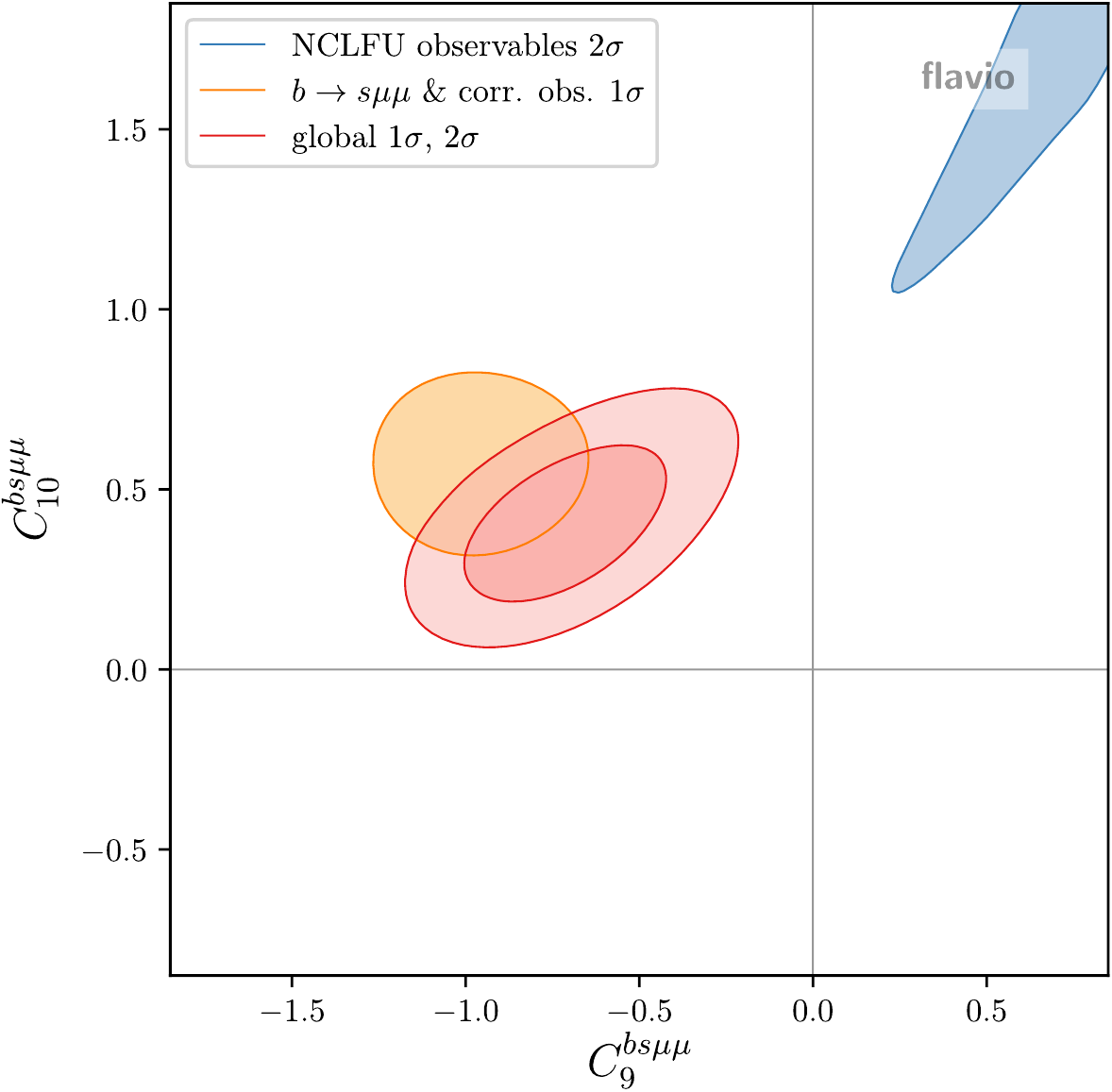}
    }
    \caption{Left panel: Measurements of the ratios $R(D)=\Gamma(B\to D\tau^- \bar{\nu})/\Gamma(B\to D\ell^-
    \bar{\nu})$ and $R(D^*)=\Gamma(B\to D^*\tau^- \bar{\nu})/\Gamma(B\to D^*\ell^- \bar{\nu})$, where $\ell$ denotes an electron or
    muon, exceed the Standard-Model predictions with a combined significance of approximately 3$\sigma$. From Ref.~\cite{HFLAVRD}.
    Right panel: a global fit of experimental data for decays of bottom hadrons to strange hadrons and charged leptons, which uses both
    lattice and non-lattice calculations of hadronic matrix elements, gives nonzero
    beyond-the-Standard-Model couplings $C_9^{bs\mu\mu}$ and $C_{10}^{bs\mu\mu}$ with a pull of approximately $6\sigma$. From Ref.~\cite{Aebischer:2019mlg} (figure simplified).}
    \label{fig:Banomalies}
\end{figure}

The CKM unitarity fit shown in Fig.~\ref{fig:unitaritytriangle} only includes the processes that are well suited to constrain a
specific side length or angle of the triangle.
There are, however, other types of decays of $b$ quarks that currently show significant deviations from the Standard Model.
One class of decays showing such deviations involves the transition of a bottom quark to a charm quark, tau lepton, and
neutrino~\cite{Ciezarek:2017yzh,HFLAVRD}, as summarized in Fig.~\ref{fig:Banomalies} (left).
The second class of decays with deviations observed involves the loop-induced (in the Standard Model) transition of a bottom quark to a strange quark and a pair of muons
or electrons~\cite{Archilli:2017xmu,Aebischer:2019mlg}, as shown in Fig.~\ref{fig:Banomalies} (right).
In both cases, the experimental results seem to suggest that the different types of leptons in nature (electrons, muons, and taus)
do not interact in a universal way as predicted by the Standard Model.
If it can be confirmed that these deviations are not caused by errors in the experimental measurements and/or theoretical calculations the implications will
be profound (many theorists are already constructing models of new fundamental physics to explain the observations; see, e.g.,
Ref.~\cite{Buttazzo:2017ixm}).
The Belle II and LHCb experiments will provide much-higher-precision measurements of the processes analyzed here and will also
measure new processes sensitive to the same underlying short-distance physics.
Both the Standard-Model predictions and the fits used to constrain new couplings beyond the Standard Model depend critically on
lattice QCD calculations of hadronic matrix elements, which need to be improved as discussed in the following sections.

\subsection{Opportunities for lattice QCD}

\subsubsection{Charm and bottom meson leptonic and semileptonic decays}
\label{sec:DBmesons}

In the charm and bottom sectors, the most precise determinations of the magnitudes $|V_{q q'}|$ of the CKM matrix elements utilize
either purely leptonic decays to a $\ell \bar{\nu}$ final state, where $\ell$ is a charged lepton and $\bar{\nu}$ is a neutrino, or
semileptonic decays to a $H \ell \bar{\nu}$ final state where in addition a hadron $H$ is present.
Up to small QED corrections, the lattice QCD inputs needed to extract $|V_{q q'}|$ from the measured decay rates are decay constants
(for the purely leptonic decays) or form factors (for the semileptonic decays).
The latter are functions of $q^2$, where $q$ is the four-momentum transfer between the initial and final hadron.
The following discussion will not go into much technical detail; we also refer the reader to the reviews of the Flavor Lattice
Averaging Group (FLAG)~\cite{FLAG}.

For the decay constants of the charmed and bottom mesons $D$, $D_s$, $B$, and $B_s$, a recent lattice QCD calculation performed by the
Fermilab Lattice and MILC Collaborations, both part of USQCD, has achieved remarkably small uncertainties of approximately $0.2\%$ ($D$,
$D_s$) and $0.7\%$ ($B$, $B_s$)~\cite{Bazavov:2017lyh}.
This is made possible by the availability of ultrafine lattice ensembles \cite{Bazavov:2012xda} and highly improved lattice discretizations, which allow the heavy quarks to be
implemented in the same way as the light quarks, eliminating a previously dominant systematic uncertainty associated with
renormalization and matching.
To reduce the uncertainty in the predicted decay rates even further, structure-dependent QED corrections need to be calculated, which
requires more complicated matrix elements from lattice QCD with an elaborate treatment of divergences associated with low-momentum
photons.
A first such calculation was recently performed for light mesons~\cite{Giusti:2017dwk,DiCarlo:2019thl}, and it would be desirable to adapt these 
techniques to heavy-meson decays. 

In the Standard Model, the purely leptonic branching ratios are suppressed by the square of the lepton mass.
As a result, the decays $B_c \to e \nu$, $B_c \to \mu \nu$, $B \to e \nu$, and $B \to \mu \nu$ have not yet been observed.
The decay $B \to \tau \nu$ has been seen, but the immediate secondary decay of the tau lepton in $B \to \tau \nu$ introduces
additional experimental challenges, and the branching fraction presently still has a large uncertainty.
Even though Belle II will make substantial improvements for these decays, the preferred method for extracting $|V_{ub}|$ and
$|V_{cb}|$ is via the semileptonic decays $B\to \pi \ell \bar{\nu}$, $B\to D \ell \bar{\nu}$, and $B\to D^* \ell \bar{\nu}$ (the
$\pi$ and $D$ mesons have spin 0, while the $D^*$ has spin 1).
For both $|V_{ub}|$ (see Fig.~\ref{fig:Vubhistory}) and $|V_{cb}|$, there are tensions between the determinations from these
``exclusive'' decay modes using form factors from lattice QCD~\cite{Harrison:2017fmw,Bailey:2014tva,Lattice:2015rga,Na:2015kha,Flynn:2015mha,Lattice:2015tia}, and inclusive determinations which involve a sum over final states that enables a theoretical
description in continuum QCD (a discussion of lattice QCD prospects for inclusive decays can be found in Sec.~\ref{sec:Luscher}).
In the case of $|V_{cb}|$, the tension is primarily driven by $B\to D^* \ell \bar{\nu}$ for which the published lattice QCD
calculations so far were restricted to the zero-recoil point~\cite{Bailey:2014tva,Harrison:2017fmw}.
There is presently an active debate whether a particular method of extrapolating the experimental data to this point, which
implements constraints based on calculations in continuum QCD and heavy-quark effective theory, is responsible for the tensions
\cite{Bigi:2017njr,Bigi:2017jbd,Bernlochner:2017jka,Bernlochner:2017xyx,Abdesselam:2018nnh,Ricciardi:2018bgs,Bernlochner:2019ldg,Dey:2019bgc,Gambino:2019sif}.
Lattice QCD calculations of the $B \to D^*$ form factors at nonzero recoil are urgently needed to settle this issue and are already
in progress by a number of groups~\cite{Aviles-Casco:2017nge,Vaquero:2019ary,Flynn:2016vej,Kaneko:2018mcr}.

The present experimental uncertainty in $|V_{cb}|$ from exclusive semileptonic $B$ decays is approximately $2\%$
\cite{Abdesselam:2018nnh}, while the theoretical uncertainty from lattice QCD is $1.4\%$~\cite{Harrison:2017fmw,Bailey:2014tva,Lattice:2015rga,Na:2015kha}.
With 5 $\rm{ab}^{-1}$ of data at Belle II, the experimental uncertainty will be reduced to $1.8\%$, and with 50 $\rm{ab}^{-1}$ of
data, the experimental uncertainty will be around $1.4\%$~\cite{Kou:2018nap}.
Although the current theory uncertainty is commensurate with the expected experimental uncertainty with 50 $\rm{ab}^{-1}$ at Belle
II, lattice calculations at nonzero recoil will help leverage the experimental data in the $q^2$ range where they are more precise. Hence they will decrease the total uncertainty in addition 
to answering the questions regarding the robustness of the extrapolation mentioned above.
For the exclusive semileptonic decay mode $B\to \pi\ell\nu$, the current experimental uncertainty is $2.5\%$~\cite{Amhis:2016xyh}
and the current lattice QCD uncertainty is $3\%$~\cite{Lattice:2015tia}.
The experimental uncertainty will be reduced to around $1.2\%$ after 50 $\rm{ab}^{-1}$ of Belle II running~\cite{Kou:2018nap}, so it is
important to improve the lattice uncertainty concurrently.
This will require high statistics, multiple fine lattice spacings, and a method that reduces or eliminates the renormalization/matching uncertainty, such as the fully relativistic treatment already used for the $B_{(s)}$-meson decay constants.
The ratio $|V_{ub}/V_{cb}|$ can also be measured by LHCb, where decay modes not involving a pion are favorable
\cite{Adinolfi:2012qfa}.
This has already been done using baryons (cf.~Sec.~\ref{sec:baryondecays}), and measurements of the ratio of $B_s \to K \mu
\bar{\nu}$ and $B_s \to D_s \mu \bar{\nu}$ decay rates are in progress~\cite{Bediaga:2018lhg}, requiring $B_s \to K$ and $B_s \to
D_s$ form factors from lattice QCD.
First calculations are already available~\cite{Bailey:2012rr,Flynn:2015mha,Monahan:2018lzv,Bazavov:2019aom,McLean:2019qcx}.

As shown in Fig.~\ref{fig:Banomalies} (left), measurements of the ratios
$R(D)=\Gamma(B\to D\tau^-\bar{\nu})/\Gamma(B\to D\ell^-\bar{\nu})$ and $R(D^*)=\Gamma(B\to D^*\tau^- \bar{\nu})/\Gamma(B\to
D^*\ell^- \bar{\nu})$, where $\ell$ denotes an electron or muon, presently exceed the Standard-Model predictions with a combined
significance of approximately 3$\sigma$~\cite{HFLAVRD}.
The experimental uncertainties on $R(D)$ and $R(D^*)$ are around $9\%$ and $5\%$, respectively.
These uncertainties will be cut in half with 5 $\rm{ab}^{-1}$ at Belle II, and will be further reduced to $3\%$ and $2\%$ with 50
$\rm{ab}^{-1}$ of data~\cite{Kou:2018nap}.
The theory uncertainties are both currently estimated at around $1\%$, but only $R(D)$ has been calculated using lattice QCD
\cite{Lattice:2015rga,Na:2015kha}.
Since $R(D^*)$ is driving the tension, it would be a good cross-check to have that ratio from the lattice as well, which again
requires the $B \to D^*$ form factors at nonzero recoil.
Moreover, the LHCb Collaboration will measure related ratios involving different species of bottom hadrons, including
$R(D_s^{(*)})$, $R(J/\psi)$ and $R(\Lambda_c^{(*)})$~\cite{Bifani:2018zmi}.
The form factors for some of these decays are presently not well known and should be calculated in lattice~QCD.

While bottom decays are presently the most interesting, there is also a lot of room for improvement in lattice calculations of charm
semileptonic decays, in particular $D \to \pi \ell \nu$ and $D \to K \ell \nu$, which can be used to extract the CKM matrix elements
$|V_{cd}|$ and $|V_{cs}|$.
In contrast to the purely leptonic charm decays, these processes have not received enough attention from the lattice community
during the past few years, but it should be possible to achieve sub-percent precision for the relevant form factors in the entire
kinematic range.

The processes discussed above are all charged-current decays, which in the Standard Model are mediated by a single $W$ boson
exchange at leading order.
A potentially greater sensitivity to physics beyond the Standard Model is provided by flavor-changing neutral-current processes such
as $B_{(s)} \to \ell^+\ell^-$ and $B \to K \ell^+\ell^-$, which in the Standard Model only occur through loops with additional virtual
particles.
The low-energy description of these processes involves a larger set of operators, of which some contribute to the decay rates
through local matrix elements, while the others contribute via nonlocal matrix elements with an additional insertion of the quark
electromagnetic current.
The local matrix elements are described by decay constants or form factors just as discussed above and are straightforward to
compute with lattice QCD, at least for the case of single, stable hadrons; see Sec.~\ref{sec:vectormesons} for a discussion of
$B\to K^*$ form factors. 
Indeed, the Standard Model predictions for rare leptonic decay $B_{(s)} \to \ell^+\ell^-$ are already very precise due (in part) to the small uncertainties in lattice QCD calculations of the $B_{(s)}$ decay constants.
In the case of rare semileptonic decays, higher-precision lattice QCD calculations of, for example, the $B \to K$ form factors (especially at large $K$ momentum, where the
uncertainty is still large) would help reduce the overall uncertainties in the fits used to extract the new-physics couplings from
the experimental data (cf.~Fig.~\ref{fig:Banomalies}, right panel).\footnote{Note that even the muon-versus-electron ratios such as
$R_K$~\cite{Aaij:2014ora} become significantly dependent on hadronic matrix elements in the presence of new physics that violates
lepton-flavor universality.} The nonlocal matrix elements, especially those involving operators with charm quarks in
$b\to s\ell^+\ell^-$ decays, are important but are very difficult to compute with lattice QCD due to the necessity to use imaginary time.
Analogous nonlocal matrix elements in rare kaon decays, where the situation is more favorable, can already be computed on the
lattice, as discussed in Sec.~\ref{sec:kaonmixingrare}. 
The new ideas discussed in Sec.~\ref{sec:Luscher} for developing lattice methods to calculate quantities involving multihadron intermediate states, if successful, could also open the door for lattice calculations of these nonlocal matrix elements.

\subsubsection{Bottom baryon decays}
\label{sec:baryondecays}

Approximately 20\% of all bottom hadrons produced at the LHC are $\Lambda_b$ baryons, and their weak decays can provide new
information on important quantities in flavor physics.

A measurement of a ratio of $\Lambda_b \to p\, \mu^-\bar{\nu}_\mu$ and $\Lambda_b \to \Lambda_c\, \mu^-\bar{\nu}_\mu$ decay rates at
LHCb, combined with a lattice QCD calculation of the $\Lambda_b\to p$ and $\Lambda_b \to \Lambda_c$ form factors has allowed the
first determination of $|V_{ub}/V_{cb}|$ at a hadron collider~\cite{Detmold:2015aaa,Aaij:2015bfa}.
The baryonic decays are chosen over the more conventional $B \to \pi \mu^- \bar{\nu}$ and $B \to D \mu^- \bar{\nu}$
decays because, with the LHCb detector, final states containing protons are easier to identify than final states with pions.
Because the baryonic decays are sensitive to both the vector and axial-vector currents in the weak effective Hamiltonian, this
measurement also disfavors right-handed couplings beyond the Standard Model as a possible explanation of the exclusive-inclusive
discrepancy in $|V_{ub}|$~\cite{Aaij:2015bfa}.
The uncertainties in $|V_{ub}/V_{cb}|$ are approximately 5\% from experiment and approximately 5\% from lattice QCD.
With LHC Run 4 data, and future higher-precision measurements of the normalization branching fraction $\mathcal{B}(\Lambda_c \to p
K^-\pi^+)$ at BESIII and Belle II, the experimental uncertainty is expected to drop below 2\%~\cite{Albrecht:2017odf}, and
commensurate improvements in the $\Lambda_b\to p$ and $\Lambda_b \to \Lambda_c$ form factors from lattice QCD are needed.

For the flavor-changing neutral current decay $\Lambda_b \to \Lambda \mu^+\mu^-$, the uncertainties from the lattice QCD calculation
of the form factors at low to moderate hadronic recoil~\cite{Detmold:2016pkz} are presently much smaller than the experimental
uncertainties~\cite{Aaij:2015xza}, but higher-precision calculations are needed to reduce the uncertainties at large hadronic
recoil.
With more precise experimental data (already expected soon from LHC Run 2), this decay will provide stringent new constraints on the
$b \to s \mu^+ \mu^-$ Wilson coefficients, where mesonic measurements currently indicate a significant deviation from the Standard
Model.
The baryonic decay $\Lambda_b \to \Lambda(\to p \pi) \mu^+\mu^-$ combines the best aspects of two different mesonic decays: $B \to
K\mu^+\mu^-$ (like the $K$, the $\Lambda$ is QCD-stable, which makes the lattice-QCD calculation easier) and $B \to K^*(\to K\pi)
\mu^+\mu^-$ (like the $K^*$, the $\Lambda$ has nonzero spin, which provides sensitivity to all Dirac structures in the weak
effective Hamiltonian).

Returning to charged-current decays, a measurement of the lepton-flavor-universality ratio $R(\Lambda_c)=\Gamma(\Lambda_b \to
\Lambda_c\,\tau^-\bar{\nu})/\Gamma(\Lambda_b \to \Lambda_c\,\mu^-\bar{\nu})$ by the LHCb collaboration is highly desired in light of
the tension seen in $R(D^{(*)})$ and is expected to be released soon.
The $\Lambda_b \to \Lambda_c\,\tau^-\bar{\nu}$ decay provides excellent sensitivity to all possible Dirac structures
\cite{Datta:2017aue}.
The current Standard Model prediction of the ratio $R(\Lambda_c)$ using the lattice QCD form factors has a $3.1\%$ uncertainty~\cite{Detmold:2015aaa},
while the projected experimental uncertainty at the end of LHC Run 4 is 1\%~\cite{Bifani:2018zmi}.
The LHCb Collaboration is also planning to measure the ratios $R(\Lambda_c^*)$, where $\Lambda_c^*$ denotes either the
$\Lambda_c^*(2595)$ with $J^P=\frac12^-$ or the $\Lambda_c^*(2625)$ with $J^P=\frac32^-$.
These ratios are expected to have smaller systematic uncertainties due to reduced feed down from higher states~\cite{LHCbRLcSt}.
Lattice QCD calculations of the relevant $\Lambda_b \to \Lambda_c^*$ form factors are therefore needed.

\subsubsection{\texorpdfstring{$B^0$-$\bar{B}^0$ and $D^0$-$\bar{D}^0$ mixing}{B0-\bar{B}0 and D0-\bar{D}0 mixing}}

The mixing of neutral $B$-mesons can be used as a powerful constraint on the CKM matrix with the mass differences of the mass
eigenstates of the $B^0_d$ and $B^0_s$ systems measured at the sub-percent level~\cite{Amhis:2016xyh}.
In addition, new physics that may be responsible for the observed tensions in $b\to s \mu^+\mu^-$ decays [Fig.~\ref{fig:Banomalies}
(right)] typically also contributes to $B_{(s)}$ meson mixing, which could provide strong constraints~\cite{DiLuzio:2018wch}.
The constraints are, however, currently limited by the theoretical errors on the hadronic mixing parameters, which are calculated in lattice QCD with uncertainties that are still an order of magnitude larger than experiment \cite{Bazavov:2016nty,Dowdall:2019bea}, leaving much room for improvement.
On the other hand, as is well known, the ratio of the mass differences benefits from error cancellations that result in a significantly smaller theoretical uncertainty, and this ratio is therefore one of the strongest constraints on the CKM unitarity triangle analysis. 
Even so, the current lattice theory error on the ratio is around $1.5\%$~\cite{Bazavov:2016nty,Boyle:2018knm,Dowdall:2019bea}, while the experimental uncertainty on this quantity is about  $0.4\%$~\cite{Amhis:2016xyh}.
Thus, improvements in these lattice-QCD calculations are necessary to fully exploit the known experimental results for their new physics discovery potential.

Most of the important, dominant sources of error in lattice calculations of $B$-mixing parameters can be greatly reduced, if not eliminated, by using the latest generations of ensembles with physical light-quark masses, small lattice spacings, and highly improved actions. In particular, employing a fully relativistic action for the $b$ quark would make it easier to adopt an entirely nonperturbative renormalization and matching procedure \cite{Carrasco:2013zta}. This is important, because perturbative truncation effects would otherwise limit the precision of a lattice calculation of $B$-mixing parameters on modern ensembles~\cite{Bazavov:2016nty,Dowdall:2019bea}.
The nonperturbative renormalizations should, in principle, be calculable with sufficient precision to yield $B$-mixing parameters with total uncertainties close to those already achieved in Ref.~\cite{Bazavov:2017lyh} for $B$-meson decay constants, i.e., at or close to the sub-percent level for the bag parameters and less than half-percent level for ratios. Further precision improvements would require the inclusion of structure-dependent QED effects. Fortunately, QED corrections to neutral $B$-mixing parameters are relatively straightforward to calculate with methods similar to those already developed for hadron masses and the hadronic vacuum polarization (see Sec.~\ref{sec:HVP}).
Given the plans for Belle II~\cite{Kou:2018nap} and LHCb~\cite{Bediaga:2012uyd,Albrecht:2017odf,Bediaga:2018lhg}, we also expect that the experimental measurements could be improved further, especially if theoretical progress were to make such an effort justifiable. 

Another observable in $B^0_s$-$\bar{B}^0_s$ mixing is the lifetime difference $\Delta \Gamma_s$.
The theory uncertainty in $\Delta \Gamma_s$ is currently much larger than the experimental uncertainty and is dominated by the
poorly known dimension-7 matrix elements.
First lattice QCD calculations of the needed matrix elements have already been started~\cite{Davies:2017jbi},
and these quantities deserve further study.

Complementary constraints on the CKM matrix and new physics can be obtained from neutral $D$-meson mixing.
Standard Model contributions to this process are dominated by the down and strange quarks, so that CP violation is strongly
suppressed; searches for CP violation in $D$-meson mixing can thus be a sensitive probe for such contributions from BSM physics.
From a theoretical standpoint, the CP-violating contributions to $D$-meson mixing in the Standard Model are much more precisely
known, since they are dominated by local $\Delta C = 2$ matrix elements as opposed to ``long-distance'' $\Delta C = 1$ processes
which are difficult to estimate and which dominate the overall mixing process.

Study of $D$-meson mixing is experimentally challenging, and current uncertainties~\cite{Amhis:2016xyh} are as large as tens of
percent, so that existing lattice-QCD calculations of the local $\Delta C=2$ matrix elements
\cite{Carrasco:2014uya,Carrasco:2015pra,Bazavov:2017weg} with errors in the 5--10\% range are sufficient for interpretation of
current experimental results in searching for CP-violating contributions.
However, future prospects for experimental measurement of $D$-meson mixing point to significant improvements, for example, LHCb
projects roughly an order of magnitude reduction in error with its Phase-II upgrade and 300 fb${}^{-1}$ of integrated luminosity
\cite{Aaij:2244311}, so more precise lattice matrix elements for $\Delta C = 2$ processes will be needed over the longer term.
A first-principles lattice calculation of the long-distance $\Delta C = 1$ contributions would also be very useful and interesting,
as theoretical uncertainties from other techniques currently used to estimate these contributions are quite large even compared to
current experimental errors.
Such a calculation would require treatment of multihadron intermediate states, which is discussed in Sec.~\ref{sec:Luscher}.

\subsubsection{Weak decays to unstable vector mesons}
\label{sec:vectormesons}

The transitions $B^0\to K^{*0}(\to K^+\pi^-) \ell^+\ell^-$, where the $K^{*0}$ is a vector meson decaying through the strong
interaction, provide particularly powerful constraints in the global fits that currently hint at new physics in the $b\to s\, \ell^+
\ell^-$ Wilson coefficients~\cite{Altmannshofer:2017yso,Capdevila:2017bsm}.
This is due to the large number of observed events and the large set of angular observables associated with the four-body final
state.
However, the only published unquenched lattice QCD calculation of the $B\to K^*$ form factors~\cite{Horgan:2013hoa}
neglects the strong decay of the $K^*$, leading to uncontrolled and unquantified systematic uncertainties.
To avoid this source of error, and to also provide information on the $K\pi$-invariant-mass distribution that goes beyond the $K^*$
resonance-pole contribution, lattice QCD calculations of $B \to K \pi$ form factors are needed.
The necessary finite-volume formalism has been developed~\cite{Agadjanov:2014kha,Briceno:2014uqa,Agadjanov:2016fbd} and is discussed in more detail in
Sec.~\ref{sec:Luscher}.
For a given angular-momentum partial wave of the $K \pi$ system (the $K^*$ resonance occurs in the $P$ wave), the $B \to K\pi$ form
factors are functions of $q^2$ (the square of the four-momentum transfer between the $B$ and the $K \pi$ system) and $s$ (the $K\pi$
invariant mass).
While the accessible range in $q^2$ is limited only by discretization and statistical errors, the accessible range in $s$ is limited
by the requirement that only two-body channels contribute to the $K\pi$ scattering.
First lattice-QCD calculations of the $B \to K \pi$ form factors at heavier-than physical up and down quark masses are underway \cite{Rendon:2018fem}.
The task for the future will be to reach the physical quark masses, and to reach few-percent precision.

A similar process involving an unstable vector meson is the decay $B \to \rho (\to\pi\pi)\ell \bar{\nu}$.
This decay can not only provide a new determination of $|V_{ub}|$ in the Standard Model, but can also put
stringent constraints on possible right-handed $b\to u$ currents beyond the Standard Model~\cite{Bernlochner:2014ova}.
The BaBar and Belle Collaborations already have data for this decay, and even more precise results are expected from the Belle~II
experiment~\cite{Kou:2018nap}.
Again, first lattice-QCD calculations using the new finite-volume formalism are underway.
One important question is what range of $s$ can be accessed when going to the physical pion mass, where the four-pion channel may
already become relevant near the $\rho$ resonance region.

\subsubsection{\texorpdfstring{$K \to \pi\pi$ decays}{K to pi pi decays}} 

Direct CP violation was measured in $K\to\pi\pi$ decays more than 15 years ago~\cite{Batley:2002gn,AlaviHarati:2002ye}, while
theorists have been trying to compute the tiny violation from first principles since the 1970s.
In 2015 the RBC/UKQCD collaborations reported that its value, $\mathrm{Re}(\epsilon'/\epsilon)$, computed in the Standard
Model is 2.1 standard deviations below Nature~\cite{Bai:2015nea}.
This is especially fertile ground for discovering new physics since there is a single complex phase in the CKM matrix that describes
all CP violation in the Standard Model.
The phase has been measured precisely in $B$ decays, which means all other instances of CP violation, like
$\mathrm{Re}(\epsilon^\prime/\epsilon)$, are tightly constrained.

The difficulty in calculating the value precisely in the Standard Model stems from the hadronic matrix elements of effective weak
interaction four-quark operators that mediate the decays between kaon and two-pion states.
The Lellouch-L\"uscher formalism needed to compute $1\to 2$ matrix elements on the lattice is discussed in Sec.~\ref{sec:Luscher}.
While the formalism is well understood, there are also numerical challenges, in particular the so-called disconnected diagrams
associated with the isospin-zero two-pion states.
The $2.1\sigma$ difference mentioned above results from roughly equal parts statistical and systematic errors in the lattice
calculation which is undergoing significant improvement.
The statistics of the original calculation, which uses special G-parity boundary conditions~\cite{Wiese:1991ku}, is being quadrupled
to cut the statistical error in half.
These special boundary conditions are needed to forbid an otherwise unphysical state where the pions are at rest (rather than
on-shell at the center of mass energy of the kaon) from being the ground state in the computed correlation function for the decay
amplitude.
However, G-parity introduces features that effectively double the cost over conventional periodic boundary conditions, and make the
calculations technically more difficult.
An alternate method, using periodic boundary conditions, is being developed by the same group to understand if state of the art
techniques can be used to extract the physical amplitude from the first excited state of the $K\to\pi\pi$ correlation function.
A successful attempt will mean  less demanding computations to address finite volume and lattice spacing systematic errors
and, at the same time, provide an important test of the G-parity method. 
Recent computations of $I=0$ pion-pion scattering which suffer the same problems appear promising.

If improved theory results for $\mathrm{Re}(\epsilon^\prime/\epsilon)$, expected in the near future, signal new physics, there are
several interesting beyond the Standard Model scenarios that can shed light on its nature (see~\cite{Buras:2018wmb} for an up to
date discussion).
An important point is that in typical BSM models either QCD or EW penguin operators dominate, but not both~\cite{Buras:2015jaq}.
So we would be in the enviable position of computing amplitudes where large cancelations are absent, unlike in the Standard Model.
BSM operators with similar Dirac structures could also explain deviations in $\epsilon_K$ and $\Delta M_K$ if current lattice
calculations suggest BSM physics there too.
The matrix elements of the local four-fermion operators relevant for $K^0$-$\bar K^0$ mixing have been computed by several lattice groups
recently~\cite{Carrasco:2015pra,Jang:2015sla,Garron:2016mva,Boyle:2017skn}, and the Standard Model bag parameter is therefore known with close to percent level precision.

\subsubsection{\texorpdfstring{$\Delta M_K$, $\epsilon_K$, and rare kaon decays}{Delta M(K), epsilon(K), and rare kaon decays}}
\label{sec:kaonmixingrare}

Especially promising phenomena in which physics beyond the Standard Model may be discovered are those that are forbidden by the
selection rules of the Standard Model, for example those in which strangeness changes by two units or a semileptonic process in
which strangeness changes but the hadronic charge does not, processes referred to as ``strangeness-changing neutral currents".
Such processes do arise in the Standard Model at second order in the weak interactions,
but they are typically suppressed by five to ten orders of magnitude, opening an important window in which beyond the
Standard Model physics might be found.

To provide meaningful tests of the Standard Model, the size of these second-order Standard Model predictions must be known.
In some cases, such as the $\Delta S=2$ indirect CP violation amplitude $\epsilon_K$ and the strangeness changing neutral current
process $K^+\to\pi^+\nu\bar{\nu}$, this second-order physics is dominated by the contribution from short distances where QCD
perturbation theory can be used.
In these cases the nonperturbative part of the calculation is the evaluation of the matrix element of a two- or four-quark operator.
For the case of $\epsilon_K$ this is the amplitude $B_K$, a $K^0$-$\bar{K}^0$ matrix element of a four quark operator which is
now known at the 1\% level from lattice calculations.
For $K^+\to\pi^+\nu\bar{\nu}$ the needed matrix element can be determined from $K_{\ell3}$ decay.
While the known, short-distance contribution to these processes is large, there are still so called ``long-distance'' contributions
which may be as large as 5\% and, especially in the case of $\epsilon_K$ will soon be needed to match the experimental precision.
For quantities such as the mass difference $\Delta M_K$ between the long- and short-lived neutral $K$ meson or the rare decay
$K_L\to\mu^+\mu^-$ such nonperturbative, long-distance contributions dominate the process and must first be determined if a search
for new physics is to be possible.

In these long-distance contributions the two $W^\pm$-$W^\mp$ or $W^\pm$-$Z^0$ exchanges can each be represented by products of two
four-quark or two-quark-two-lepton operators that are separated by a distance large compared to $1/M_W$, typically by the Compton
wavelength of the charm quark or larger and nonperturbative methods are needed for their evaluation.
The use of lattice gauge theory to compute these quantities has been developed over the past eight years and is
on a solid theoretical footing~\cite{Christ:2012se,Bai:2014cva,Christ:2015aha,Christ:2016eae,Christ:2016mmq,Bai:2017fkh,Bai:2018hqu}.

These are challenging calculations at the frontier of what is currently possible with lattice QCD.
The required lattice amplitudes are complex and computationally expensive four-point functions.
As a second-order calculation in effective field theory, new counterterms may appear which come from short-distance effects.
These have already been computed in QCD and electro-weak perturbation theory~\cite{Buchalla:1995vs} but need to be more precisely
determined.
There are potentially important finite-volume effects which can be computed and removed~\cite{Christ:2015pwa}.
States with energies below the kaon mass will contribute unphysical terms which grow exponentially with the time separations present
in the Green functions being computed and which must be independently computed and subtracted.

At present, exploratory calculations for all but the $K_L\to\mu^+\mu^-$ process have been undertaken and more advanced calculations
with physical parameters are planned or underway.
The most mature is a calculation of $\Delta M_K$~\cite{Bai:2018mdv} at physical quark masses using 152 gauge-field
configurations on a $64^3\times128$ lattice.
The results from this calculation have approximately 25\% statistical errors and systematic discretization errors caused by the inclusion of
the heavy charm quark (to realize the GIM mechanism) that need to be investigated further in future calculations.
RBC/UKQCD expects to obtain a result for $\Delta M_K$ with a controlled 20\% total error within five years.
A calculation for the decay $K^+\to\pi^+\nu\bar{\nu}$, which has been started, will aid in the interpretation of the results of the NA62 experiment at CERN.

A large-scale lattice QCD study of the process $K\to\pi\ell^+\ell^-$ by RBC/UKQCD is now also underway.
This process can be viewed as an electromagnetic correction to the usual nonleptonic weak transition $K\to\pi$ which is of interest
because of current hints of $\mu$-$e$ universality violations.
This calculation, which integrates out the charm quark perturbatively, is expected to yield first results in 2020, with a calculation
which includes the charm quark expected in the future.
While a first calculation of the long-distance contribution to $\epsilon_K$~\cite{Christ:2015phf} has been performed, further study
with physical quark masses will be begun when the needed personnel are available and the calculation of $\Delta M_K$ better
understood.
A result for this important 5\% effect on $\epsilon_K$ with 20\% error may be expected in 5-7 years.
Finally a numerical strategy for the valuation of the combined second-order electromagnetic and first order weak process
$K_L\to\mu^+\mu^-$ is currently being developed with an exploratory calculation perhaps two years away.

It should be emphasized that such calculations of rare kaon decays and $\epsilon_K$ are needed if on-going experiments are to
realize their full potential to reveal physics beyond the Standard Model.
The calculation of both $\Delta M_K$ and $K_L\to\mu^+\mu^-$ will allow legacy measurements from KTeV to
become sensitive tests of the Standard Model, with a discovery potential created by these lattice-QCD calculations equivalent to that
of new large-scale experiments such as NA62 at CERN or KOTO at JPARC.

\subsubsection{Multihadron physics}
\label{sec:Luscher}

Lattice-QCD calculations provide matrix elements between finite-volume states.
For a single-particle state, finite-volume effects are exponentially suppressed and numerically very small, whereas for two or more
particles the effects fall like inverse powers of the box size $L$ and must be accounted for.
Furthermore, a finite-volume state necessarily includes components of all particle combinations that are allowed by kinematics and
strong interaction selection rules.
For example, a finite-volume state with $I=Q=S=0$ and $E \approx M_D$ will consist of two-pion, four-pion, six-pion, $K \bar K$,
$\eta\eta$ and other components, within each of which there will be contributions from multiple relative angular momenta.
This state is thus very different from the in- or out-states that enter into infinite-volume matrix elements, which contain a single
component, e.g.
two pions.
The theoretical challenge is to relate these two types of matrix element.
This has been achieved for processes involving multiple two-particle channels~\cite{He:2005ey,Lage:2009zv,Bernard:2010fp,Briceno:2012yi,Hansen:2012tf,Briceno:2014uqa,Briceno:2015csa,Briceno:2015tza,Baroni:2018iau}, based on the seminal work of
L\"uscher~\cite{Luscher:1986pf,Luscher:1991cf} and Lellouch \cite{Lellouch:2000pv}, and has been implemented in several lattice
calculations~\cite{Dudek:2014qha,Wilson:2014cna,Dudek:2016cru,Moir:2016srx,Briceno:2016kkp,Briceno:2017qmb,Alexandrou:2018jbt}.

The L\"uscher method involves two steps.
First, a quantization condition is derived, which describes the finite-volume energy spectrum.
In the two-particle case this depends on infinite-volume scattering amplitudes within and between channels, as well as known
kinematic functions that depend on the box size and shape.
Lattice results for the spectrum at several box sizes and total three-momenta can then be used to determine the suitably
parametrized scattering amplitudes.
The second step is to relate matrix elements involving the finite-volume states to those involving a single measured in- or
out-state.
In the single-channel two-particle case this involves only a normalization constant, the Lellouch-L\"uscher factor, which can
be determined from the energy dependence of the scattering amplitude~\cite{Lellouch:2000pv}.
For multiple two-particle channels one requires a linear combination of matrix elements, as well as information on the
energy dependence of the $S$~matrix~\cite{Hansen:2012tf,Briceno:2014uqa,Briceno:2015csa,Briceno:2015tza}.
While this is more complicated, all required quantities can be calculated with 
lattice QCD.

We now turn to the prospects for calculations involving states with three or more particles.
Examples are the $K\to 3\pi$ decays, hadronic $D$ decays, inclusive $B$ decays, and neutral $D$-meson mixing.

To study these using lattice QCD, additional formalism is
needed. Given the promise of realistic lattice calculations, this kind of mathematical physics is an active
area of research.
Two approaches are being followed: adding particles one at a time (3 particles, then 4 particles, etc.), and directly determining
the shape of the inclusive amplitudes (corresponding to summing over any number of particles).
We discuss these two approaches in turn.

The generalization to three particles has been the focus of much effort in the last five years, and considerable progress has been
made, following the pioneering work of Refs.~\cite{Polejaeva:2012ut,Briceno:2012rv}.
The first step, namely the derivation of a quantization condition, has been achieved for a relativistic system of three identical
particles~\cite{Hansen:2014eka,Hansen:2015zga}, including the possibility of mixing with states containing two
particles~\cite{Briceno:2017tce} and resonant subprocesses \cite{Briceno:2018aml}. Several important cross-checks of the formalism have been carried out~\cite{Meissner:2014dea,Hansen:2015zta,Hansen:2016fzj,Hansen:2016ync,Meng:2017jgx,Sharpe:2017jej,Pang:2019dfe}. A recent review can be found in Ref.~\cite{Hansen:2019nir}.
A simpler form has been derived for the three-particle case using a nonrelativistic effective field
theory~\cite{Hammer:2017uqm,Hammer:2017kms,Pang:2019dfe}, while other relativistic approaches have also been
considered~\cite{Mai:2017bge,Mai:2018djl}.
Finite-volume energies now depend not only on two-particle scattering amplitudes but also on three-to-three amplitudes.
Furthermore, the formalism of Ref.~\cite{Hansen:2014eka} provides a parametrization of the infinite-volume three-particle scattering amplitude that is unitary~\cite{Briceno:2019muc}, and has been shown to be equivalent to parametrizations used to analyze experimental scattering data~\cite{Jackura:2019bmu}.
Using simple parametrizations of the two- and three-particle amplitudes, as well as other well-motivated approximations, the quantization condition has been
solved in simple examples~\cite{Hammer:2017uqm,Briceno:2018mlh,Doring:2018xxx,Mai:2018djl,Blanton:2019igq,Romero-Lopez:2019qrt}.
Because this formalism is needed to study most resonances in QCD---a topic of great interest in hadronic physics---it is likely that
in the next few years a workable form of the three-particle quantization condition will be developed, including generalizations to
nondegenerate particles and particles with spin.
This can then be applied to the cases of interest to flavor physics, e.g., the $3\pi$ state.

There has, to date, been no work on the generalization of the Lellouch-L\"uscher factor to the three-particle
case. This is needed, for example, to use lattice simulations to determine the
$K\to3 \pi$ amplitude. However, we expect that the approach of Ref.~\cite{Briceno:2014uqa}
will be generalizable to three particles, and that the requisite Lellouch-L\"uscher relation will be developed
in parallel with the maturation of the three-particle quantization condition.

For applications to hadronic $D$ decays, in which charm-decay CP violation was recently observed for the first time \cite{Aaij:2019kcg},
further developments are needed. This is because,
as noted above, states with $E=M_D$ involve not only two particles but also four pions
(and perhaps also six pions). Thus the L\"uscher approach needs to be extended further.
This is certainly possible in principle; experience with implementing the three-particle case
will be useful in determining whether a practical methodology can be developed.
This is work on a five-year timescale.
The extension to four  particles is also essential for a lattice study of neutral $D$ mixing,
because there are significant long-distance contributions
involving four-pion intermediate states in the Standard Model.

With the L\"uscher approach, extracting all relevant energy levels on the lattice is a significant computational challenge,
especially when channels with more than two particles are involved, and when the lattice volume is large, resulting in a dense
spectrum.
The progress in excited-state multihadron spectroscopy is discussed further in the companion whitepaper
``Hadrons and Nuclei''~\cite{Detmold:2018qcd}.

In certain cases it is also possible to extract inclusive observables from the lattice, i.e.~quantities that are summed over all
multiparticle final states.
This is relevant for inclusive $D$ and $B$ decay rates, for $D$ and $B$ neutral meson mixing, and for semileptonic decays that are
inclusive with respect to the hadronic final states.
The latter would, for example, be very helpful for inclusive $|V_{ub}|$ determinations that suffer from a large $b \to c \ell
\bar{\nu}$ background (see, e.g., Ref.~\cite{TheBABAR:2016lja}).

To study inclusive quantities, one must again address the issue that lattice calculations are performed in a finite volume and can
only access imaginary-time correlation functions.
It turns out that the restriction to imaginary time presents no challenge for creating single particle states, so that it is
possible to calculate finite-volume matrix elements such $\langle D|\mathcal{H}_w(\tau, \bm{x}) \mathcal{H}_w(0)|D\rangle_L$, where
$\mathcal{H}_w$ is an insertion of the weak Hamiltonian density, allowing the $D$-meson state, $\vert D \rangle$, to decay.
However, this still does not directly give access to the observable of interest, since $\mathcal{H}_w(\tau, \bm{x})$ depends on the
imaginary time coordinate $\tau$.
If it were instead defined with a Minkowski 4-vector, $x$, and in an infinite volume, this matrix element would directly give the
total decay rate of $D$ to all final states coupled by $\mathcal{H}_w$.
This is achieved simply by Fourier transforming $x = (t, \bm{x})$ to zero spatial momentum and zero injected energy.
Inserting a complete set of states between the two weak Hamiltonians then leads to the desired expression: a sum over rates,
$\vert\langle E = M_D, \alpha, \text{out}|\mathcal{H}_w|D \rangle|^2$, running over all open channels, with standard phase space
integrals to include all final-state kinematics.

The challenge is thus to understand whether this information can be recovered from the finite-volume observable
$\langle D|\mathcal{H}_w(\tau, \bm{x})\mathcal{H}_w(0)|D \rangle_L$, where the subscript $L$ indicates the lattice size,
and $\tau=it$ is Euclidean time.
Just as one can relate the real time coordinate, $t$, to energy via a Fourier transform, to relate $\tau$ to the physical energy,
one must solve the inverse Laplace transform~\cite{Liu:1993cv,Hansen:2017mnd}.
This is numerically very challenging but is also a universal problem across many branches of physics and related fields.
For this reason a large body of work has gone into developing algorithms and understanding limitations.
(See, for example, Ref.~\cite{Press:2007zz}.) %
These ideas are already being applied in nonzero-temperature lattice QCD (see, for example,
Refs.~\cite{Asakawa:2000tr,Meyer:2011gj}), and are expected to provide an important new tool in extracting decays and transitions
involving many multiparticle states.

One specific example is the Backus-Gilbert approach, developed by geophysicists Backus and Gilbert in the late 1960s
\cite{Backus:1968,Backus:1970}.
In the context of $D$ decays, the method gives a linear mapping from $\langle D|\mathcal{H}_w(\tau, \bm{x}) \mathcal{H}_w(0)
\vert D \rangle_L$ to a ``smeared out'' version of the total rate, in which the final state energy is integrated over some window
about the target energy, $M_D$~\cite{Liu:2016djw,Hansen:2017mnd}.
By extracting an observable with limited energy resolution, the severity of the inverse problem is reduced so that the target
observable does not suffer from significantly enhanced uncertainties.
In fact, the Backus-Gilbert algorithm uses the correlation matrix from the input data and, in this way, designs an optimal inverse
that strikes a balance between energy resolution and uncertainties in the final result.

Remarkably, the imperfect energy resolution also serves to reduce finite-volume effects that would otherwise dominate these types of
multihadron inclusive observables.
In this way the Backus-Gilbert approach solves two problems simultaneously: by targeting an observable with finite energy
resolution, the difficulty of the inverse Laplace transform is reduced and the finite-volume effects are
suppressed~\cite{Hansen:2017mnd}.
Although the idea is new, numerical investigations already show promising results~\cite{Liang:2017mye,Liang:2019frk}.
Future calculations will aim to extrapolate to infinite volume and, through improved data, achieve increasingly fine energy
resolution.
This will open a window to a large class of multihadron observables, in which all multiparticle states are included without the
need to disentangle each contribution individually.

As mentioned above, in addition to inclusive hadronic and semileptonic decay rates, this approach could potentially be used to study
the dominant, long-distance contributions to $D_0$-$\bar{D}_0$ mixing.
This is again given by an inverse Laplace transform of $\langle D|\mathcal{H}_w(x) \mathcal{H}_w(0)|D \rangle$,
although in this case for all final state energies rather than just $E=M_D$.
Convoluting this result with a kernel, roughly given by $1/(E-M_D)$, then provides the target long-distance mixing observables.
It is important to note that this idea is preliminary so that future work is required to test the feasibility in lattice-QCD
calculations.
If successful this will mean a set of new and novel observables for the lattice.

Here we have focused on one of many algorithms for inverting the Laplace transform.
A powerful alternative is the maximum-entropy method (MEM)~\cite{Jaynes:1957zza,Jaynes:1957zz}.
Like the Backus-Gilbert algorithm, this gives an estimate of the desired inverse, but is instead driven by minimizing a smoothness
function.
The approach is commonly employed in astronomical synthesis imaging, where the resolution depends on the signal-to-noise ratio,
which must be specified~\cite{Thompson:2001}.
Thus it has an analogous smoothening characteristic that can again be used to suppress finite-volume effects.
In practice both methods, as well as other techniques, will be applied to extract the available information for this novel class of
observables.

Finally, we note that another lattice QCD approach to inclusive semileptonic $B$ decays has recently been proposed in
Ref.~\cite{Hashimoto:2017wqo}, which does not involve solving the inverse problem.
The strategy is instead to use Cauchy's integral formula to analytically continue the amplitude from the experimentally accessible
physical kinematic region to an unphysical region in which the lattice calculation can be performed.
To perform the Cauchy integral, the experimental data need to be reanalyzed to provide a differential distribution in the relevant
variable(s).

It is important to appreciate that the mathematical aspects of the research described in this subsection, by its nature, leads
to pilot computing projects to demonstrate their feasibility.
Access to computing clusters with rapid turnaround have been essential for the impressive progress of the past few years, and such
computing facilities will continue to be key to developing the tools for multihadron quark-flavor physics.

\section{Lepton flavor physics}
\label{sec:LF}
\subsection{Experimental motivation}

For more than ten years, the {approximately 3.5 to 4.0 standard-deviation
difference between the experimentally measured value of the muon's anomalous magnetic moment,
$a_\mu=(g_\mu-2)/2$, and the one calculated in the Standard Model has
provided an experimentally viable and theoretically plausible
possibility of discovering new physics.  The significance of this
discrepancy has been stable since the measurement by experiment E821
at Brookhaven National Laboratory~\cite{Bennett:2006fi} and originates
in approximately equal parts from experimental and theoretical
uncertainties.

Experiment E989 at Fermi National Accelerator Laboratory, running
since March 2018, seeks to reduce the total error on $a_\mu$ achieved
by E821 (0.54 ppm) by a factor of four.  Experiment E989 uses the same
strategy -- and even the same muon storage ring -- as the earlier
E821.  A second, complementary, experiment to measure $a_\mu$ using
ultracold muons is being mounted at J-PARC. This experiment, E34,
plans to begin taking data around 2020.  Clearly, the theory
uncertainty on $a_\mu$ must be reduced to fully utilize the improved
experimental measurements.  In Sec.~\ref{sec:leptonopp}, we present a
plan for reducing the theoretical errors, which stem primarily from
nonperturbative hadronic contributions, to the needed level of
precision using lattice QCD.  This work is already underway, and much
progress has been made towards reaching this goal.

Already, however, the Fermilab E989 experiment has collected more than
the total statistics of the Brookhaven E821 experiment.  
The announcement of a first result for $a_\mu$ based on this data set is
anticipated in the second half of 2019.  On this shorter timescale, one may
be able to obtain a more precise determination of $a_\mu$ in the
Standard Model by combining results from analyses of
$e^+e^-$-scattering data and independent lattice-QCD calculations.
Various hybrid lattice-plus-$R$-ratio approaches for reducing the
theory error are being explored, and one such example is outlined in
Sec.~\ref{sec:leptonopp}.

If the present discrepancy between the experimental measurement and
theory for the muon anomalous magnetic moment is indeed due to
beyond-the-Standard-Model physics, one expects the new particles or
interactions to produce effects in other, related observables.
Interestingly, a new measurement of the fine-structure-constant in
Cesium~\cite{Parker191} with a three-times smaller uncertainty than
the previous determination from Rb increases the significance of the
difference between theory and experiment for the electron anomalous
magnetic moment to $2.4\sigma$~\cite{Jegerlehner:2018zrj}.  The
difference in $a_e$ is opposite in sign to that in $a_\mu$, but is
consistent with some new physics scenarios \cite{Davoudiasl:2018fbb}.
Although the significance of the disagreement is, at present,
dominated by the uncertainties on the experimental
measurements~\cite{PhysRevLett.100.120801,Hanneke:2010au} further
reduction of the experimental errors is anticipated \cite{PrivateCommBerkeleyNorthwestern},
at which point more precise theory will be required.  As
described in Sec.~\ref{sec:leptonopp}, lattice QCD can be employed to
calculate the hadronic contributions to $a_e$ in the same manner as
for $a_\mu$.  Lattice methods are also being developed that would
enable the first full, independent cross-check of the five-loop QED
calculation of $a_e$~\cite{Aoyama:2017uqe}.

\subsection{Opportunities for lattice QCD}
\label{sec:leptonopp}

The current difference between the experimental (EXP) measurement and the Standard Model (SM)
prediction of the muon anomalous magnetic moment is
\begin{align}
  a_\mu^{\rm EXP} - a_\mu^{\rm SM} = 27.4\underbrace{(2.7)}_{\rm HVP}\underbrace{(2.6)}_{\rm HLbL}
      \underbrace{(0.1)}_{\rm other~SM} \underbrace{(6.3)}_{\rm EXP} \times 10^{-10},
\end{align}
where the dominant theoretical uncertainties from the hadronic vacuum polarization contribution (HVP)~\cite{Blum:2018mom}, the hadronic light-by-light
contribution (HLbL)~\cite{Prades:2009tw}, and the BNL E821 experiment~\cite{Bennett:2006fi} are given separately.
With the anticipated reduction of the experimental uncertainty to approximately $1.6 \times 10^{-10}$ by 2020, a reduction
of the uncertainties on both the HVP and HLbL contributions to a commensurate level will be needed to fully benefit from the experimental investment.

The calculation of $a_\mu$ in the Standard Model begins with the quantum-field-theory matrix element of the electromagnetic current between on-shell muon states,
\begin{equation}
    \langle \mu(p') | J_\rho(q) |\mu(p)\rangle = ie\bar u(p') \left(\gamma_\rho F_1(a^2) + 
        \frac{i \sigma_{\rho\nu}q^\nu}{2m_\mu} F_2(q^2)\right)u(p),
\label{eq:amp}
\end{equation}
describing the physical interaction of a muon with an external field (photon).
When the momentum transferred to the muon by the field ($q=p-p'$) goes to zero, the value of the form factor $F_2(0) = a_\mu$.
At tree level, a nonrelativistic expansion of the Dirac equation yields $F_2(0)=0$, so $a_\mu$ can only be nonzero through radiative corrections.
These can be computed order-by-order in perturbation theory in~$\alpha$.  The first term in the expansion, arising at one-loop order, was computed 
by Schwinger over 60 years ago, and is simply $\alpha/2\pi=0.001\,161\,714\, 9\cdots$.
The leading hadronic terms enter at two- and three-loop order in $\alpha$, respectively, and are suppressed by the relevant hadronic scales
$\Lambda_{\rm QCD}$ as $m_\mu^2 / \Lambda_{\rm QCD}^2$.  These hadronic corrections are therefore roughly four and six orders of magnitude smaller than the Schwinger term.
Even so, the uncertainties on the HVP and HLbL contributions to $a_\mu$ are sufficiently large that they dominate the theory.

In Secs.~\ref{sec:HVP} and~\ref{sec:HLbL}, we outline concrete
opportunities for lattice QCD to reduce both the HVP and HLbL
uncertainties on this timescale.  In order to realize these
improvements, a concerted community effort is required.  This
motivated the creation of the Muon $g-2$ Theory Initiative in 2017,
whose steering committee is co-chaired by USQCD members A.\ El-Khadra
and C.\ Lehner.  This effort has held four workshops since its
inception~\cite{TI:1,TI:2,TI:3,TI:4} and will continue the workshop
series in 2019 at the Institute for Nuclear Theory in
Seattle~\cite{TI:5}.  In addition, USQCD is devoting substantial human
effort and computing resources towards the task of bringing the
hadronic uncertainties on $a_\mu$ to the level needed by the Fermilab
E989 and J-PARC E34 experiments.

For the electron anomalous magnetic moment, the tension between theory and experiment is
\begin{align}
    a_e^{\rm{EXP}} - a_e^{\rm{SM}} = -87 \underbrace{(23)}_{\alpha}\underbrace{(02)}_{\rm SM}\underbrace{(28)}_{\rm EXP}
        \times 10^{-14},
\end{align}
where the uncertainties from
experiment~\cite{PhysRevLett.100.120801,Hanneke:2010au}, the
fine-structure-constant~\cite{Parker191}, and the Standard-Model
theory calculation~\cite{Jegerlehner:2018zrj} are given separately.
The theory uncertainty stems in approximately equal parts from the
hadronic contributions and from the five-loop QED contribution
calculated in Ref.~\cite{Aoyama:2017uqe}.  Although the experimental
uncertainty on $a_e$ is presently much larger than the theory error, a
reduction of uncertainties on $a_e^{\rm{EXP}}$ and $\alpha$ by an
order of magnitude may be feasible in the next few
years~\cite{PrivateCommBerkeleyNorthwestern}.  Therefore, strategies
must be devised and methods developed to reduce the theoretical error
on $a_e^{\rm SM}$ on this time scale.  We outline how the lattice-QCD
community can contribute to this goal in Sec.~\ref{sec:HLQED}.

\subsubsection{Hadronic vacuum polarization}
\label{sec:HVP}

The HVP contribution arises from the magnetic parts of the upper diagrams shown in Fig.~\ref{fig:hadronic diagrams}.
\begin{figure}
    \centering
    \includegraphics[width=0.2\textwidth]{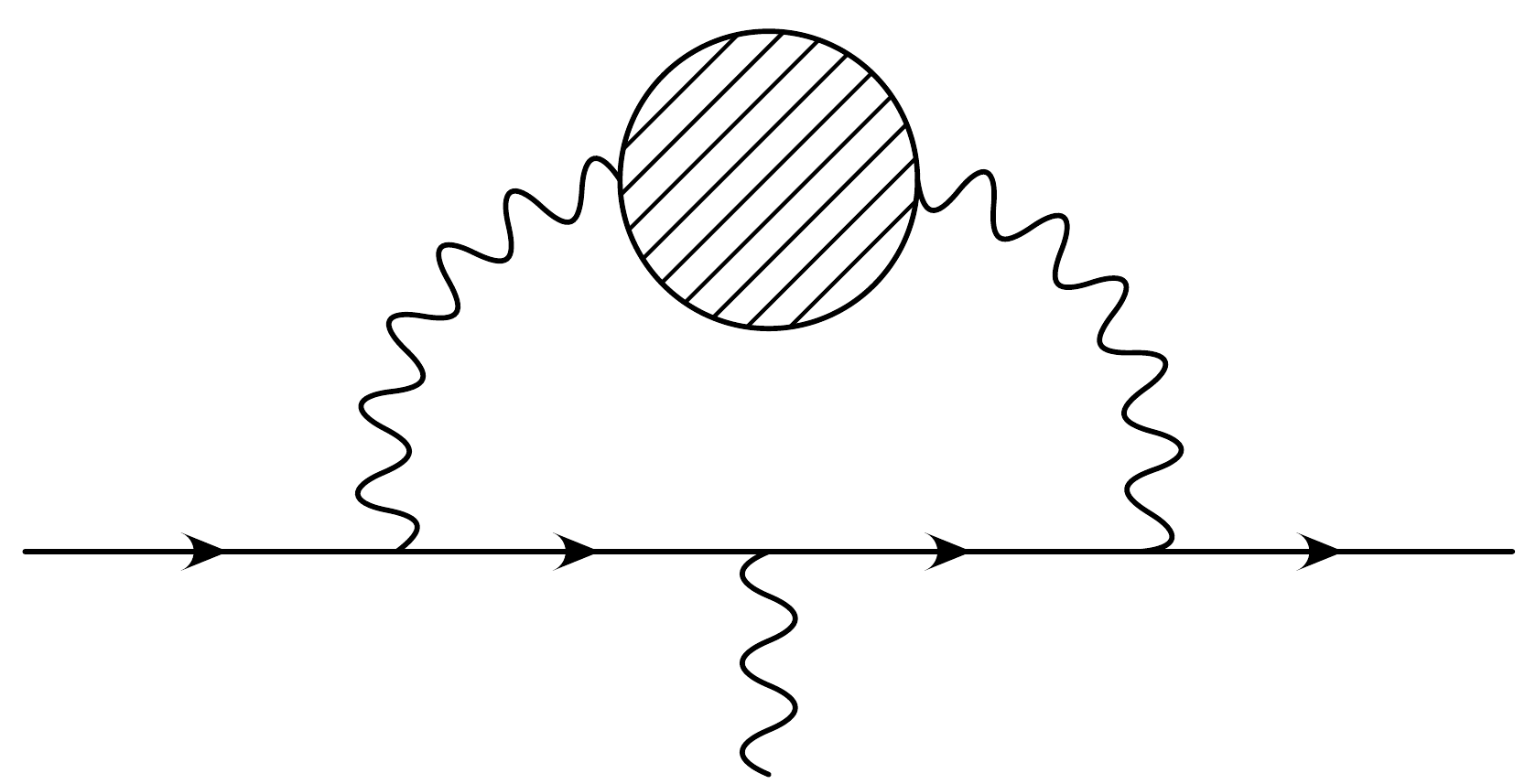}~~~~
    \includegraphics[width=0.2\textwidth]{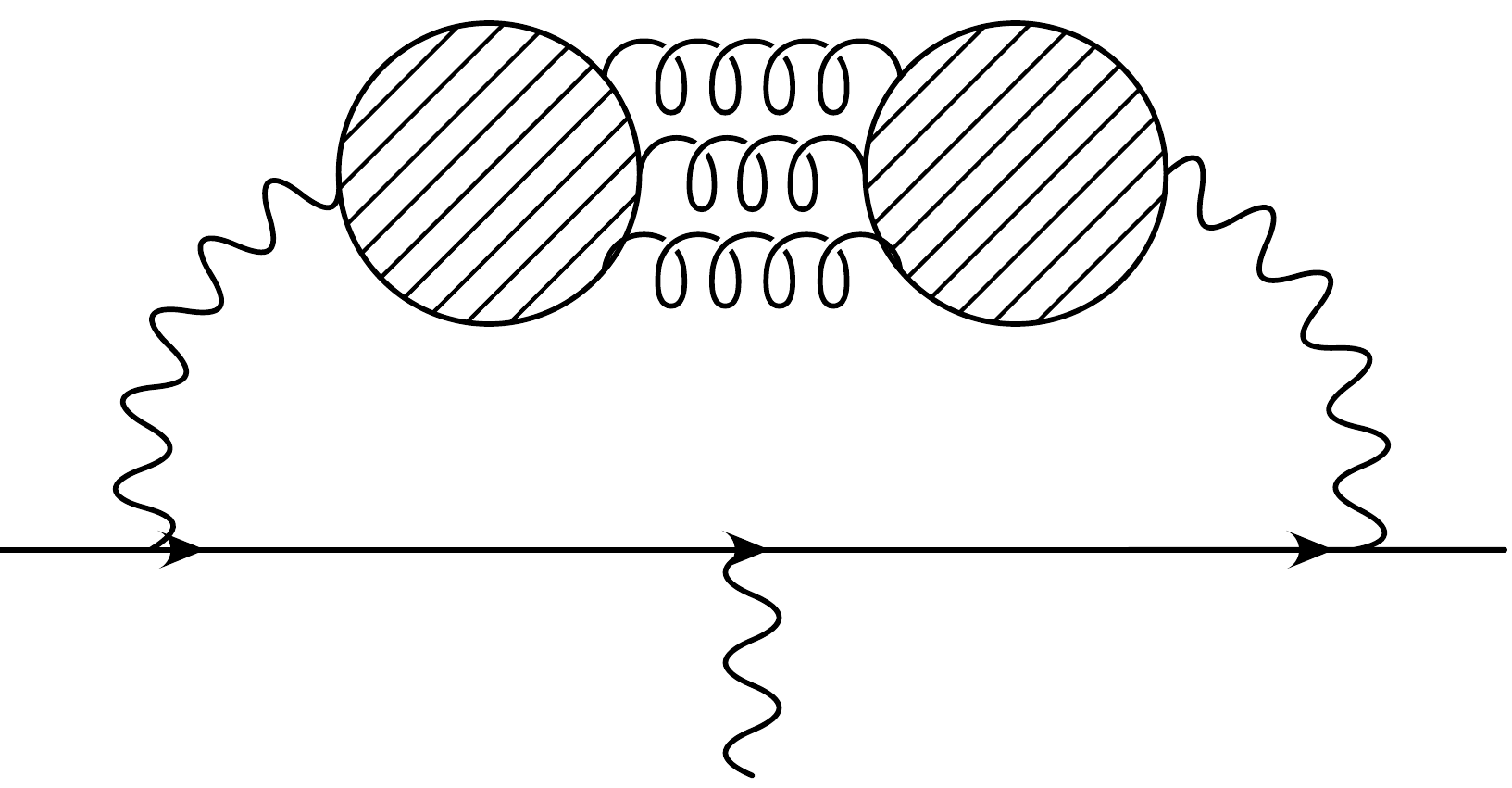}~~~~
    \includegraphics[width=0.2\textwidth]{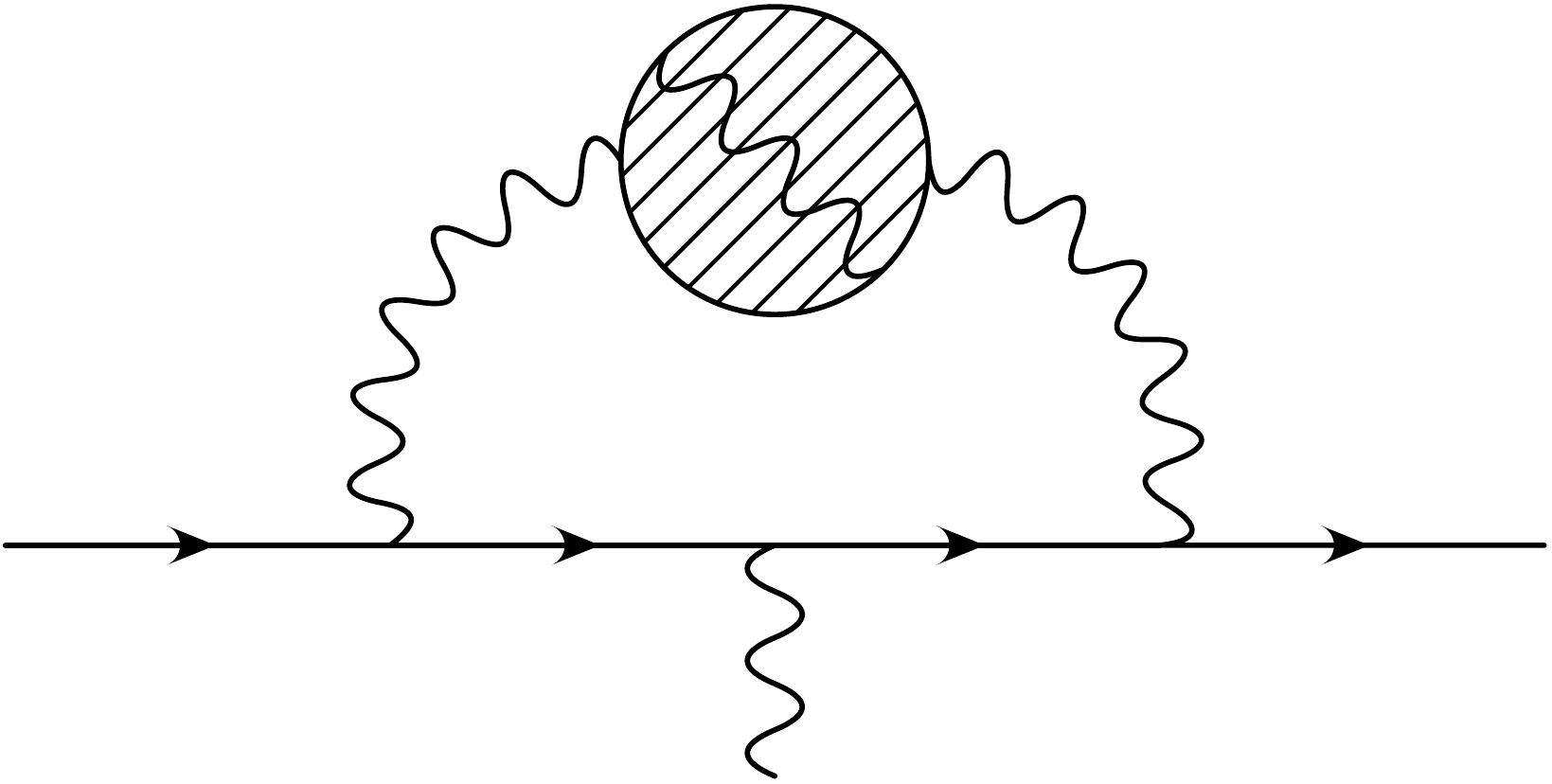}~~~~
    \includegraphics[width=0.2\textwidth]{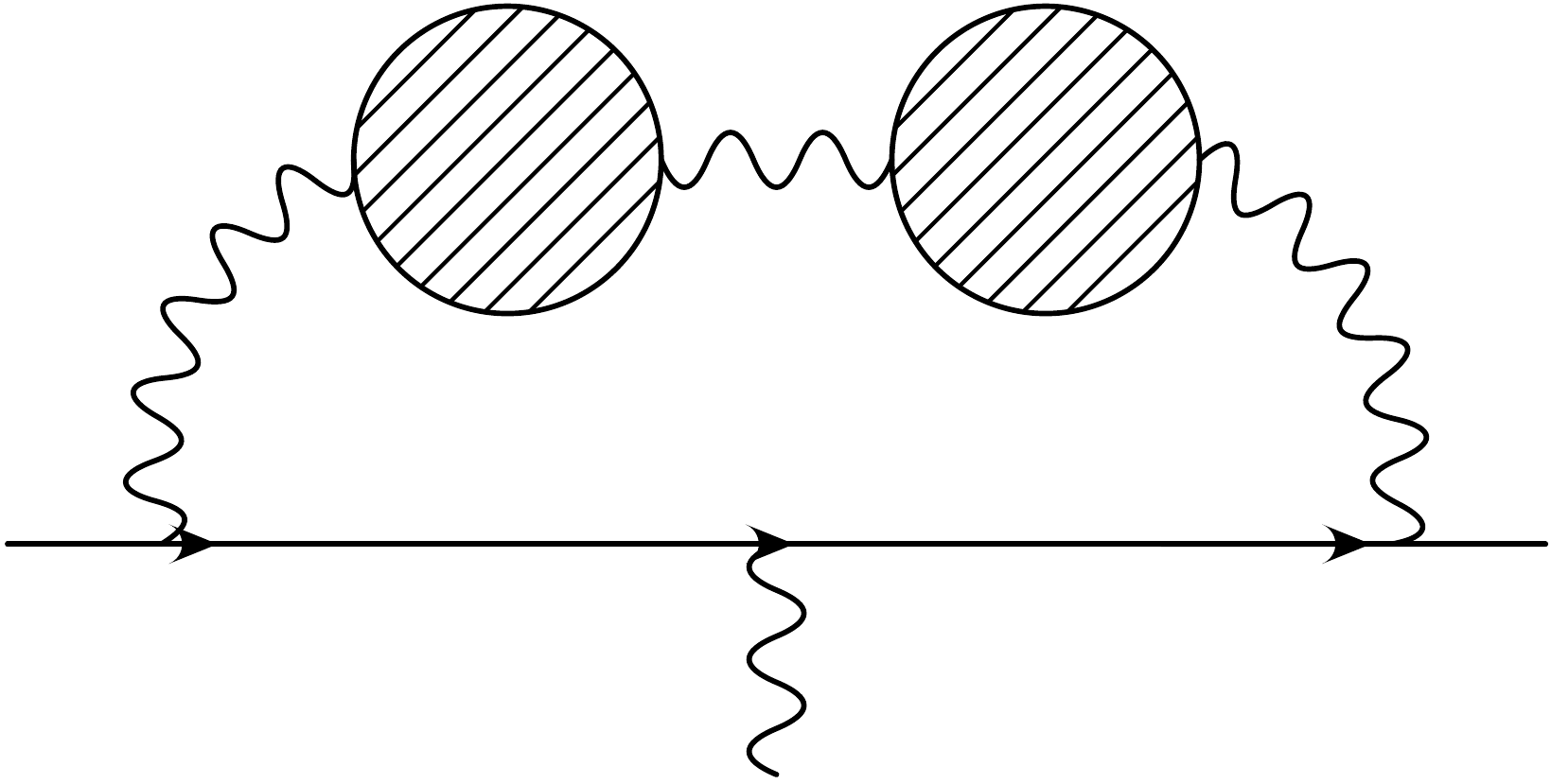}\\
    \includegraphics[width=0.2\textwidth]{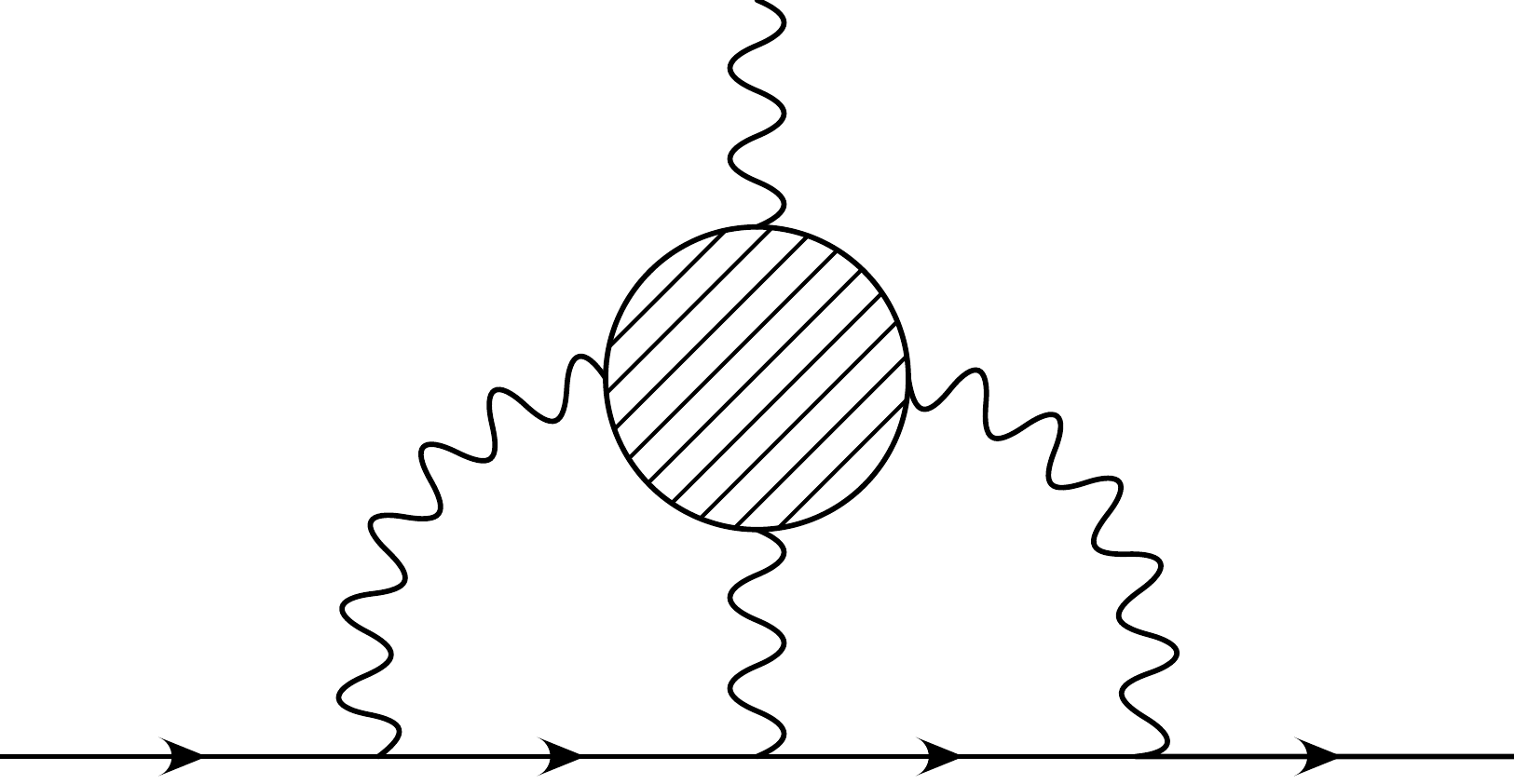}~~~~
    \includegraphics[width=0.2\textwidth]{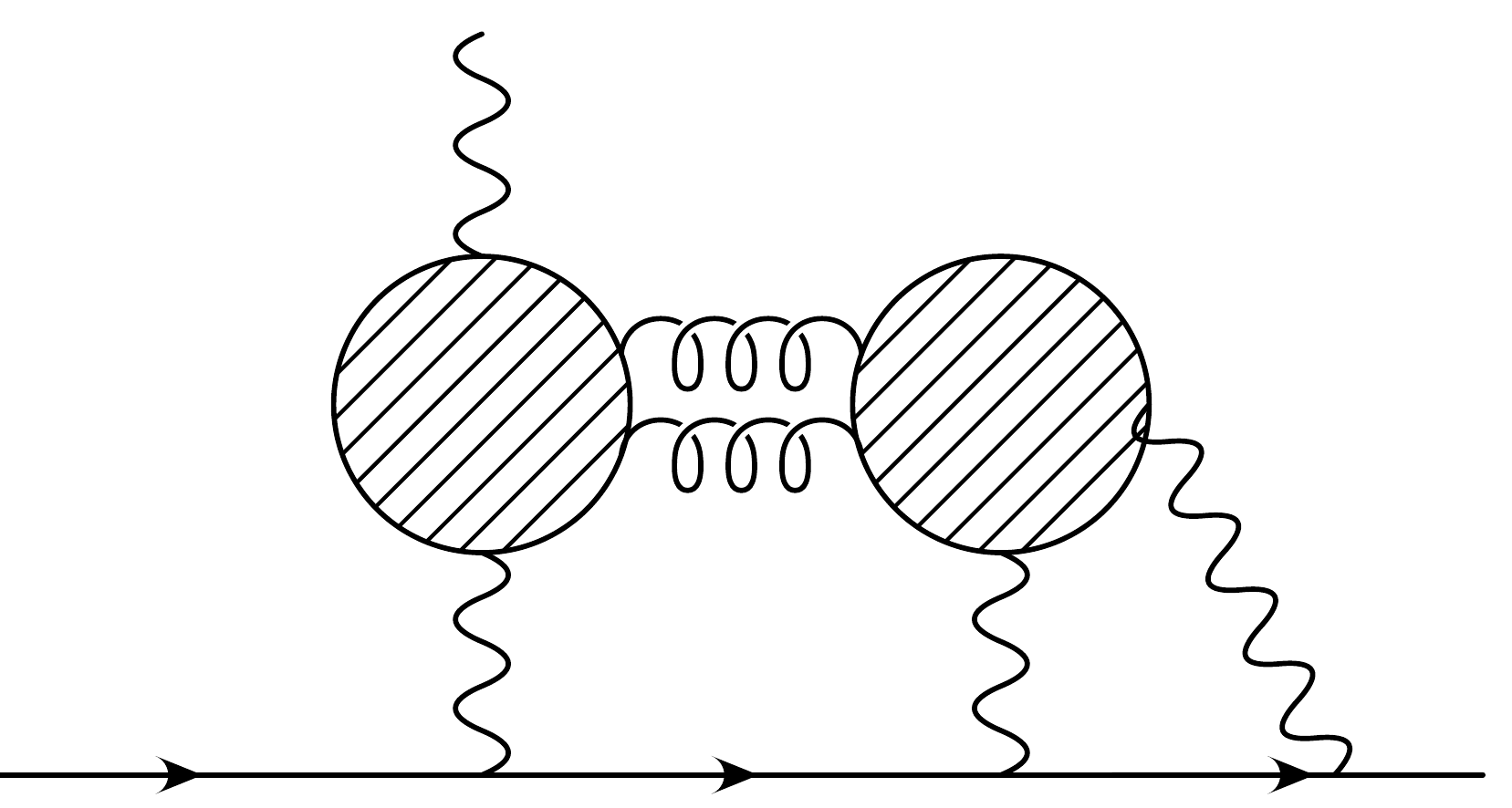}
    \caption{The upper diagrams from left to right show the leading-order (LO) quark-connected, LO quark-disconnected, LO QED
    corrections, and an example of next-to-leading order (in $\alpha$) HVP diagrams.
    The lower diagrams show the leading quark-connected (left) and quark-disconnected (right) contributions to the HLbL contribution.
    Subleading diagrams with up to four quark loops in light-by-light scattering are not shown.}
    \label{fig:hadronic diagrams}
\end{figure}
The HVP can be computed directly in lattice QCD or using a dispersion relation from the total cross section of $e^+e^-\to$ hadrons
($R$~ratio) or $\tau$ decays into hadrons and a neutrino.
Lattice gauge theory here requires the inclusion of QED to achieve high precision but, as usual,
is systematically improvable.
The dispersive method requires control of perturbative QCD and an effective description of radiative corrections, which is typically
performed in scalar QED.
In the case of $\tau$ decays, additional isospin-breaking corrections are needed.
In principle both methods can be improved beyond their current precision.
At the moment, the $R$-ratio method has the smallest uncertainty; however, in the presence of conflicting BaBar and KLOE data
sets~\cite{Davier:2017zfy,Keshavarzi:2018mgv} the common choice of inflating local uncertainties in $R(s)$ using the PDG $\chi^2$
prescription is not unique.
In order to reduce the dependence on this choice, a combined lattice and $R$-ratio analysis which removes parts of the conflicting
data sets, as suggested in Ref.~\cite{Blum:2018mom}, is valuable.
Such a combined analysis can now be performed with an uncertainty of $2.7\times 10^{-10}$, which yields a result consistent with the
currently most precise pure $R$-ratio result of Ref.~\cite{Keshavarzi:2018mgv}.

So far the lattice community has computed
connected~\cite{Burger:2013jya,Chakraborty:2016mwy,DellaMorte:2017dyu,Borsanyi:2017zdw,Blum:2018mom,Giusti:2018mdh,Davies:2019efs,Shintani:2019wai},
disconnected~\cite{Blum:2015you,Borsanyi:2017zdw,Blum:2018mom}, and isospin breaking~\cite{Chakraborty:2017tqp,Blum:2018mom,Giusti:2019xct}
contributions to the leading-order HVP.
In addition, a dedicated calculation of the next-to-leading order HVP
has recently been published in Ref.~\cite{Chakraborty:2018iyb}.
Figure~\ref{fig:hadronic diagrams} shows a diagrammatic classification of these contributions.
In Fig.~\ref{fig:hvp summary}, we list these recent results which currently approach a total uncertainty of approximately
$15\times10^{-10}$.
\begin{figure}
    \centering
    \includegraphics[width=0.7\textwidth]{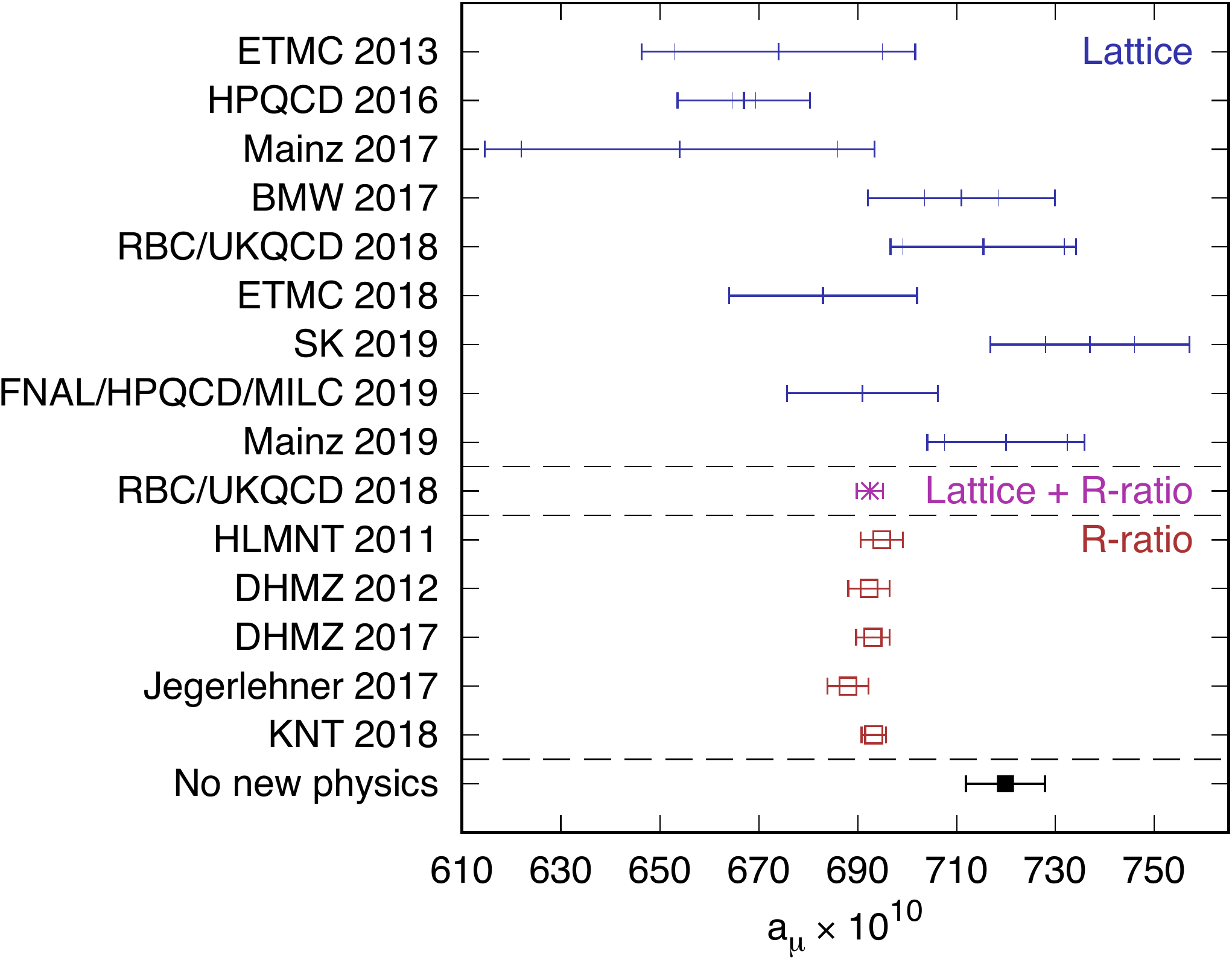}
    \caption{Leading-order HVP contributions to $a_\mu$ from recent lattice (blue) and dispersive (red) calculations.
      The purple data-point is a combined analysis of lattice and dispersive input~\cite{Blum:2018mom}.
        The referenced contributions are
            ETMC 2013~\cite{Burger:2013jya},
            HPQCD 2016~\cite{Chakraborty:2016mwy},
            Mainz 2017~\cite{DellaMorte:2017dyu},
            BMW 2017~\cite{Borsanyi:2017zdw},
            RBC/UKQCD 2018~\cite{Blum:2018mom},
            ETMC 2018~\cite{Giusti:2018mdh},
            SK 2019~\cite{Shintani:2019wai},
            FNAL/HPQCD/MILC 2019~\cite{Davies:2019efs},
            Mainz 2019~\cite{Gerardin:2019rua},
            HLMNT 2011~\cite{Hagiwara:2011af},
            DHMZ 2012~\cite{Davier:2010nc},
            DHMZ 2017~\cite{Davier:2017zfy},
            Jegerlehner 2017~\cite{Jegerlehner:2017lbd},
            KNT 2018~\cite{Keshavarzi:2018mgv},
        and ``No new physics''~\cite{PDG2018}.
        The innermost error-bar corresponds to the statistical uncertainty.}
    \label{fig:hvp summary}
\end{figure}
USQCD members have played a pioneering role in many of these contributions, such as the first calculation of strong
isospin-breaking effects at physical pion mass~\cite{Chakraborty:2017tqp}, the first calculation of QED corrections at physical pion
mass~\cite{Blum:2018mom}, as well as the first calculation of a combined lattice and $R$-ratio calculation at physical pion mass in
lattice QCD+QED~\cite{Blum:2018mom}.

Assuming adequate computing resources, there is still much the lattice
community will do to reduce errors on the HVP, perhaps by an order of
magnitude, in the next few years.  The RBC/UKQCD collaboration is
developing a distillation-based~\cite{Peardon:2009gh} method for
exclusive-channel calculations of the vector-vector correlation
function needed for the HVP at physical pion mass, see also
Refs.~\cite{Feng:2014gba,Andersen:2018mau} for earlier results at
unphysical pion mass.  This allows both for an improvement of the
bounding method~\cite{Blum:2018mom,Borsanyi:2017zdw}, which addresses
the large signal-to-noise problem in the HVP, as well as control of
finite-volume errors~\cite{Meyer:2011um}.  The distillation method
will also be used to achieve a per-mille level lattice spacing
determination, which is needed to reach the Fermilab E989 target
uncertainty.  Preliminary results on this new method have been shown
by RBC/UKQCD in recent workshops of the theory initiative
\cite{TI:2,TI:4}.  The Fermilab/HPQCD/MILC collaboration is also
actively exploring a similar multi-operator method.  Further
improvements are also expected for the QED and strong isospin breaking
corrections to the HVP with active efforts by both Fermilab/HPQCD/MILC
and RBC/UKQCD.

Finite-volume errors in the dominant quark-connected light-quark contribution are exponentially suppressed in the volume, but their
control at the needed precision poses another challenge for a pure lattice calculation to reach the Fermilab E989 target precision.
While it has been demonstrated that leading-order chiral perturbation theory underestimates the effect for typical
lattices~\cite{TI:4}, a method that adds in the resonance contribution, such as the one of Ref.~\cite{Chakraborty:2016mwy}, or a
direct determination of the effect via the elastic pion-scattering as proposed in Ref.~\cite{Meyer:2011um} may provide a viable path
to control this uncertainty.
Recently, first results have also been presented by RBC/UKQCD~\cite{TI:4} which contrast these ideas directly with lattice
computations at physical pion mass computed at different volumes.

The outlined improvements, combined with continued support for the needed computational effort, should allow for a complete
first-principles calculation of the HVP with subpercent precision in 2019 and with precision comparable to the Fermilab E989
experiment by 2022.

As already stated, also $\tau$ decay data can be used to determine the HVP contribution given that isospin-breaking effects can be
computed with sufficient precision.
First progress towards a computation of these corrections has recently been reported by RBC/UKQCD~\cite{Bruno:2018ono}.
This will add a valuable cross-check and may help resolve the tensions between $e^+e^-$ and $\tau$ determinations of the
HVP~\cite{Jegerlehner:2011ti,Davier:2017zfy,Cirigliano:2018dyk}.
In the longer term, direct measurements in the space-like region, e.g., by the proposed MUonE experiment at
CERN~\cite{Venanzoni:2018ktr} may provide additional cross-checks and improvements.
Lattice QCD can provide important input to this type of HVP determination and first results have already been
published~\cite{Giusti:2018mdh}.

The lattice computations of the HVP are a good example of the
importance of the hardware resources of the USQCD collaboration.  Many
of the methods employed to obtain the results discussed in this
section were developed and tested on these resources. In addition the
USQCD resources were used to carry out the needed computations on all
but the most demanding ensembles. Hence, these resources have been
crucial for the progress made so far and will continue to be important
as the computations are pushed to the next stage. However the
availability of leadership class facilities, and in particular the
advent of the new Summit supercomputer in Oak Ridge will also play an
important role in facilitating further improvements by allowing high
statistics lattice calculations on the most demanding ensembles, which
are needed to reduce the total uncertainties to the E989 target level.

\subsubsection{Hadronic light-by-light scattering}
\label{sec:HLbL}

The HLbL contribution, while smaller by about two orders of magnitude than the HVP, is not as well
known.
The most recent combined error for the HVP is now at the same level as the HLbL error, $2.6\times 10^{-10}$~\cite{Prades:2009tw}.
This value stems from a 2008 
workshop in Glasgow and is often referred to as the ``Glasgow consensus''.
The central value and the error stem from a set of models that cannot be systematically improved, in contrast
to first-principles lattice-QCD calculations.
So far only the RBC collaboration has published results for a lattice calculation of the HLbL
contribution~\cite{Blum:2014oka,Blum:2015gfa,Blum:2016lnc}.
The best of these to date gives $5.35\pm 1.35 \times10^{-10}$~\cite{Blum:2016lnc} compared with
$10.0\pm2.6\times10^{-10}$ from the Glasgow consensus~\cite{Prades:2009tw}.
The former was obtained for physical masses, but at a single lattice spacing and volume, and includes only the leading disconnected
contribution (lower right diagram in Fig.~\ref{fig:hadronic diagrams}).
Nevertheless, lattice QCD is already impacting the Standard Model result: it suggests the model calculations are not wildly off, at
least not enough to bring the Standard Model into agreement with experiment.

To nail down the HLbL value, the above-mentioned systematics must be controlled.
RBC is currently calculating on several lattices to enable a continuum limit, and on a series of ensembles with fixed lattice
spacing to take the infinite-volume limit.
Both systematics may be large, of order 20--50\%, based on results in pure QED~\cite{Blum:2015gfa} and preliminary results in~QCD.
For the finite-volume case, a two-pronged approach is being followed, one where the entire calculation including the QED parts is
performed in finite volume (in the QED$_{\rm L}$ formulation~\cite{Hayakawa:2008an}) and then extrapolated to infinite volume, and a
second where the QED part (essentially a two-loop integral) is done directly in infinite volume and in the
continuum~\cite{Blum:2017cer}, similar to the HVP calculations.
The latter approach was pioneered by the Mainz group~\cite{Asmussen:2016lse,Asmussen:2017bup}, who have performed calculations for HLbL scattering at
unphysical masses~\cite{Green:2015sra,Gerardin:2017ryf,Asmussen:2018oip} but have not yet combined the two into a calculation of $a_\mu$.
While the second approach eliminates the power law errors from QED, and therefore has only exponentially small errors from QCD, it
suffers from larger statistical errors.
Similarly, in QED$_{\rm L}$ it has been observed that the signal degrades appreciably as the volume grows and the massless photons
propagate to longer and longer distances.
It is not clear at this stage which method will have the smallest total error, so both are being vigorously pursued.
Thankfully the QCD part, the dominant cost of the calculations, does not have to be computed twice.
Based on this methodology and a series of calculations at different lattice spacings and volumes, RBC expects to publish the first
complete first-principles calculation with controlled errors in 2019, before first results of the Fermilab E989 experiment.

In the longer term, the uncertainties of lattice QCD can be substantially reduced by treating the dominant pion-pole
contribution separately.
This is possible in a controlled fashion and is being pursued by both the RBC and Mainz collaborations.
Such a separation of individual contributions may also allow for cross-checks against recent progress in dispersive methods such as
the results of Refs.~\cite{Hoferichter:2014vra,Colangelo:2015ama,Gerardin:2016cqj,Colangelo:2017qdm,Colangelo:2017fiz,Hoferichter:2018dmo,Hoferichter:2018kwz,Gerardin:2019vio}.

\subsubsection{High-loop QED contributions}
\label{sec:HLQED}

As outlined above, the electron anomalous magnetic moment may also
provide a promising opportunity in the next few years, when both the
experimental measurement of $a_e$ and of the fine-structure constant
have been improved by another order of magnitude.  The current
uncertainty of the Standard Model prediction of $a_e$ originates in
approximately equal parts from hadronic contributions and from the five-loop
QED calculation of Kinoshita's group~\cite{Aoyama:2017uqe}.  For improvements
of the former, the same lattice methods apply as for the muon, and it is
expected that these improvements could be made as by-products for the muon
case.

More interestingly, perhaps, is the five-loop QED calculation.  Due to its
tremendous complexity, so far there is no complete independent cross-check
of the results of Ref.~\cite{Aoyama:2017uqe}.  We believe that this should
be viewed as an opportunity for the lattice community to revive efforts towards
a stochastic high-loop determination of $a_e$ with lattice methods.  Such efforts
have been started using numerical stochastic perturbation theory~\cite{Burgio:1997fp}
and the diagrammatic Monte Carlo technique~\cite{Rappl2015}, but neither has
been brought to completion.  A dedicated effort of the lattice community
on this interesting challenge may prove useful and promises access to higher-order
results as well.

\section*{Acknowledgments}
TB, NHC, AXK, SM, and SRS acknowledge support by the United States Department of Energy, Office of Science, Office of High Energy Physics under Award Numbers D{E-S}C0010339, D{E-S}C0011941, D{E-S}C0015655, D{E-S}C0009913, and D{E-S}C0011637, respectively. AXK also acknowledges support by the Fermilab Distinguished Scholars Program. Brookhaven National Laboratory is supported by the United States Department of Energy, Office of Science, Office of High Energy Physics, under Contract No.\ D{E-S}C0012704.
Fermilab is operated by Fermi Research Alliance, LLC, under Contract No.\ DE-AC02-07CH11359 with the United States Department of Energy, Office of Science, Office of High Energy Physics.

\end{document}